\documentclass[prb,showpacs,floatfix,twocolumn]{revtex4}

\usepackage{color,amssymb,graphicx,txfonts}

\def\ZM{Zn$_{1-x}$Mg$_x$O}
\def\ZMM{Zn$_{0.875}$Mg$_{0.125}$O}

\begin{document}

\title{Local structures of polar wurtzites \ZM\/ studied by Raman and 
$^{67}$Zn/$^{25}$Mg NMR spectroscopies and by total neutron scattering} 

\author{Young-Il Kim,$^1$ Sylvian Cadars,$^2$ Ramzy Shayib,$^2$ 
Thomas Proffen,$^3$ Charles S. Feigerle,$^4$ Bradley F. Chmelka,$^2$ 
and Ram Seshadri$^{1,5}$} 
\affiliation{$^1$Materials Department and Materials Research Laboratory, 
University of California, Santa Barbara, California 93106, USA}  
\affiliation{$^2$Department of Chemical Engineering, 
University of California, Santa Barbara, California 93106, USA} 
\affiliation{$^3$Los Alamos National Laboratory, Lujan Neutron Scattering 
Center, LANSCE-12, Los Alamos, New Mexico 87545, USA} 
\affiliation{$^4$Department of Chemistry, 
The University of Tennessee, Knoxville, Tennessee 37996, USA} 
\affiliation{$^5$Department of Chemistry and Biochemistry, 
University of California, Santa Barbara, California 93106, USA} 

\date{\today} 

\begin{abstract} Local compositions and structures of \ZM\/ alloys have been 
investigated by Raman and solid-state $^{67}$Zn/$^{25}$Mg nuclear magnetic 
resonance (NMR) spectroscopies, and by neutron pair-distribution-function 
(PDF) analyses. The $E_2^\mathrm{low}$ and $E_2^\mathrm{high}$ Raman modes 
of \ZM\/ display Gaussian- and Lorentzian-type profiles, respectively. 
At higher Mg substitutions, both modes become broader, while their 
peak positions shift in opposite directions. The evolution of Raman spectra
from \ZM\/ solid solutions are discussed in terms of lattice deformation 
associated with the distinct coordination preferences of Zn and Mg. 
Solid-state magic-angle-spinning (MAS) NMR studies suggest that the 
local electronic environments of $^{67}$Zn in ZnO are only weakly 
modified by the 15\,\% substitution of Mg for Zn. 
$^{25}$Mg MAS spectra of Zn$_{0.85}$Mg$_{0.15}$O show an unusual upfield 
shift, demonstrating the prominent shielding ability of Zn in the nearby 
oxidic coordination sphere. 
Neutron PDF analyses of \ZMM\/ using a 2\,$\times$\,2\,$\times$\,1 supercell 
corresponding to Zn$_7$MgO$_8$ suggest that the mean local geometry of MgO$_4$ 
fragments concurs with previous density functional theory (DFT)-based 
structural relaxations of hexagonal wurtzite MgO. 
MgO$_4$ tetrahedra are markedly compressed along their $c$-axes and are smaller in 
volume than ZnO$_4$ units by $\approx$\,6\,\%. 
Mg atoms in \ZM\/ have a shorter bond to the $c$-axial oxygen atom than to the three 
lateral oxygen atoms, which is distinct from the coordination of Zn. 
The precise structure, both local and average, of \ZMM\/ obtained from time-of-flight 
total neutron scattering supports the view that Mg-substitution in ZnO results in 
increased total spontaneous polarization. 

\end{abstract}

\pacs{71.55.Gs, 77.22.Ej, 78.30.-j, 61.05.fm, 61.05.Qr} 

\maketitle 

\section{Introduction} 

Research in the area of polar semiconductor heterostructures has been growing
rapidly, driven in large part by interest in two-dimensional electron gas 
(2DEG) systems.\cite{Tsuka,Koike,Rajan,Sasa,Tampo} 2DEGs are known to form at 
heterojunction interfaces that bear polarization gradients.
They can display extremely high electron mobilities, especially at low
temperatures, owing to spatial confinement of carrier motions.\cite{Davies} 
Recent reports of 2DEG behaviors in Ga$_{1-x}$Al$_x$N/GaN  and 
\ZM/ZnO heterostructures have great significance for the development of 
novel high-electron-mobility transistors (HEMTs)\cite{Rajan,Sasa,Tampo} and 
quantum Hall devices.\cite{Tsuka} 

2DEG structures are usually designed by interfacing a 
polar semiconductor with its less or more polar alloys in an epitaxial 
manner. Since the quality of the 2DEG depends critically on interface 
perfection, as well as the polarization gradient at the heterojunction, 
understanding compositional and structural details of the parent 
and alloy semiconductors is an important component in 2DEG design 
and fabrication. The evolution of atomic 
positions and cell parameters upon alloying can directly affect the magnitude 
of the polarization gradient and the carrier density at the heterojunction. 

\ZM/ZnO is one of the more promising heterostructure types for studies of 
2DEGs, due to the large polarization of ZnO, the relatively small lattice 
mismatch, and the large conduction band offsets in the \ZM/ZnO heterointerface.
Although 2DEG formation in \ZM/ZnO heterostructures have been researched 
for some time, a clear understanding of the alloy structure of \ZM\/ is 
currently lacking. Recently, we have studied composition-dependent 
changes in the crystal structures of polycrystalline \ZM\/ by synchrotron 
x-ray diffraction and Raman
spectroscopy.\cite{Kim1,Kim2}  For the composition range 
$0\,\leqslant\,x\,\leqslant\,0.15$, we have shown that Mg-substitution 
modifies the aspect ratio of  the hexagonal lattice through enhanced bond 
ionicity, and in parallel, decreases static polarization in the crystal, due to
decreased internal distortion in the tetrahedral coordinations. 

Here, we conduct a detailed and more precise study of the local 
structure of \ZM\/ alloys using Raman and solid-state nuclear magnetic resonance 
(NMR), in conjunction with neutron diffraction techniques. 
Raman and NMR spectroscopy are useful probes for addressing the molecular compositions 
and structures of solid-solution systems. 
Peak shapes and widths of the Raman spectra reflect compositional 
fluctuations and both short- and long-range order, whereas NMR is sensitive to 
the local environments around specific nuclei. Thus, both techniques can 
provide structural information on \ZM\/ that is complementary to diffraction 
analyses. Here, we examine the details of $E_2^\mathrm{low}$ and 
$E_2^\mathrm{high}$ Raman modes for \ZM\/ ($x=0$, 0.05, 0.10, and 0.15). 
We have also used spin-echo magic-angle-spinning (MAS) $^{67}$Zn and $^{25}$Mg 
NMR to study separately zinc and/or magnesium species in ZnO 
and Zn$_{0.85}$Mg$_{0.15}$O. 
These measurements have been correlated with average and local crystal structures 
of \ZM\/ and ZnO, as established by 
Rietveld (diffraction space) and pair-distribution-function (PDF, real-space) 
analyses of time-of-flight neutron diffraction data. 
Compared with x-rays, neutron scattering provides much greater sensitivity 
to oxygen and Mg positions, as well as larger momentum transfer as measured
by the larger maximum $Q$ wavevector.\cite{Proffen} By taking advantage of the 
increased data quality, we are able to isolate the geometry of MgO$_4$ moieties 
stabilized in the wurtzite ZnO lattice.  

\section{Experimental}

Polycrystalline powder samples of ZnO and \ZM\/ ($x$ = 0.05, 0.10, 0.125, 
and 0.15) were prepared from oxalate precursors obtained by co-precipitation 
using Zn(CH$_3$CO$_2$)$_2\cdot$2H$_2$O, Mg(NO$_3$)$_2\cdot$6H$_2$O, and 
H$_2$C$_2$O$_4$, all of which had purities of 99.999\,\% from Aldrich. 
The two metal salts were dissolved together in deionized water and added 
to an oxalic acid solution in the ratio of 
[Zn$^{2+}$]:[Mg$^{2+}$]:[C$_2$O$_4^{2-}$] = (1$-x$):$x$:1.05. 
Upon mixing with oxalate, Zn$^{2+}$ and Mg$^{2+}$ immediately coprecipitated 
as white crystalline oxalate powders, which were washed with deionized water 
and dried at 60$^{\circ}$C for 4\,h and subsequently heated at 550$^{\circ}$C
for 20\,h in air to decompose the oxalates to the oxides. 
Powder x-ray diffraction measurements confirmed the formation of 
Zn$_{1-x}$Mg$_x$C$_2$O$_4\cdot$2H$_2$O after heating at 60$^{\circ}$C,
and \ZM\/ after decomposing at 550$^{\circ}$C. 
For the NMR experiments, samples were also prepared at different conditions of 
temperature (900$^{\circ}$C) and atmosphere (O$_2$, N$_2$). 

Raman spectra for ZnO and \ZM\/ ($x=$ 0.05, 0.10, and 0.15) were acquired at 
room temperature using a Jobin Yvon-Horiba T64000 triple grating (1800\,gr/mm) 
spectrometer. The spectra of lightly compressed powders were recorded using 
micro-Raman sampling in air with 514.5\,nm and 675.5\,nm excitations. 
The spectra reported here were averages of 10 acquisitions of 20 s 
integrations of the CCD detector with $\approx$10\,mW of laser power focused 
onto the samples through a 10$\times$ objective. There was no evidence of 
degradation of the samples or associated changes in their spectra under these conditions. 
The spectrometer was calibrated using a 520.7\,cm$^{-1}$ lattice mode 
of silicon. The Raman features of wurtzite $E_2^\mathrm{low}$ and $E_2^\mathrm{high}$ 
phonon modes were analyzed in detail to determine the peak shape, position, 
and width. Spectral background was removed following Shirley\cite{Shirley}, 
and the profile fitting was performed using pseudo-Voigt\cite{Cox} or 
Breit-Wigner-Fano functions.\cite{Yoshik}

NMR measurements of the low-gyromagnetic-ratio 
nuclei $^{67}$Zn and $^{25}$Mg were conducted at high (19.6\,Tesla) 
magnetic field strength at the National High Magnetic Field Laboratory in Tallahassee, Florida. 
These investigations benefited from the enhanced sensitivity and improved resolution 
for the $^{67}$Zn and $^{25}$Mg nuclei that results from reduced second-order 
quadrupolar interactions, which scale inversely with 
the strength of the high applied magnetic field. 
The experiments were conducted at room temperature at 19.6\,T ($^1$H resonance frequency 
of 830\,MHz), which for $^{67}$Zn ($I=\frac52$, 4.1\,\% natural abundance, 
\textit{ca.} 1.5 receptivity relative to $^{13}$C) and 
$^{25}$Mg ($I=-\frac52$, 10.1\,\% natural abundance, \textit{ca.} 0.7 
receptivity relative to $^{13}$C) corresponded to Larmor frequencies of 51.88\,MHz and 
50.76\,MHz, respectively. 
All of the NMR spectra presented here were recorded 
on a single-resonance 4-mm probehead under MAS conditions at 10\,kHz, 
using a Hahn-echo 
(\textit{i.e.} $\frac{\piup}{2}$-$\tauup$-$\piup$-$\tauup$-acquisition) 
with the delays $\tauup$ set to one rotor period $\tauup_R$. 
The $^{67}$Zn and $^{25}$Mg shifts were referenced 
to the bulk external secondary standards 
ZnSe (274\,ppm relative to 1.0\,M \textit{aq.} Zn(NO$_3$)$_2$) and 
MgO (26\,ppm relative to 3.0\,M \textit{aq.} MgSO$_4$)\cite{Dupree}, 
respectively. Pulse lengths of 2\,$\muup$s and 4\,$\muup$s were used for 
the $\frac{\piup}{2}$ and $\piup$ pulses, respectively. 
A recycle delay of 1\,s was used in each case, using 16,000 transients 
for the $^{67}$Zn MAS spectra (\textit{ca.} 5\,h each), and 160,000 transients 
for the $^{25}$Mg MAS experiments (\textit{ca.} 46\,h each). 
Second-order quadrupolar MAS lineshapes were fitted using the program 
\textsc{DMfit}.\cite{Massiot}

Time-of-flight neutron diffraction data for ZnO and \ZMM\/ were obtained 
on the neutron powder diffractometer NPDF at the Lujan Neutron Scattering 
Center at Los Alamos National Laboratory. 
For ease of structural modeling in the PDF analysis of \ZM, we chose 
a composition of $x=\frac18$ and have used an appropriate 16-atom supercell 
model of Zn$_7$MgO$_8$. 
For each sample, $\approx$\,2\,g of powder were packed in a vanadium can, 
and the data  were collected for 3\,h at 25$^{\circ}$C using four detector 
banks located  at 46$^{\circ}$, 90$^{\circ}$, 119$^{\circ}$, and 148$^{\circ}$. 
Control runs, for the intensity corrections, employed an empty vanadium can, 
a vanadium piece, and air. 
The program \textsc{PDFgetN} was used to extract the PDF $G(r)$ 
from the raw scattering data.\cite{Peterson}
First, the coherent scattering intensity $I(Q)$ was  obtained from raw data 
by the intensity corrections for container, background, and incident beam. 
Then, the $I(Q)$ was converted to the structure factor $S(Q)$ by the 
corrections for sample absorption, multiple scattering, and inelasticity. 
Finally, the PDF $G(r)$ was constructed by the Fourier transform of the reduced 
structure factor $F(Q)=Q[S(Q)-1]$. In the Fourier transform of $F(Q)$ to 
$G(r)$, $Q$ data were terminated at 35\,\AA$^{-1}$. PDF refinements were 
performed using the software program \textsc{PDFgui}.\cite{Farrow} 
The average crystal structures of ZnO and \ZMM\/ were determined by the 
Rietveld method using the \textsc{GSAS-EXPGUI} software 
suite.\cite{Larson,Toby} 

\section{Results and discussion} 

\subsection{Raman spectroscopy} 
\begin{figure} \smallskip 
\centering \includegraphics[width=8cm]{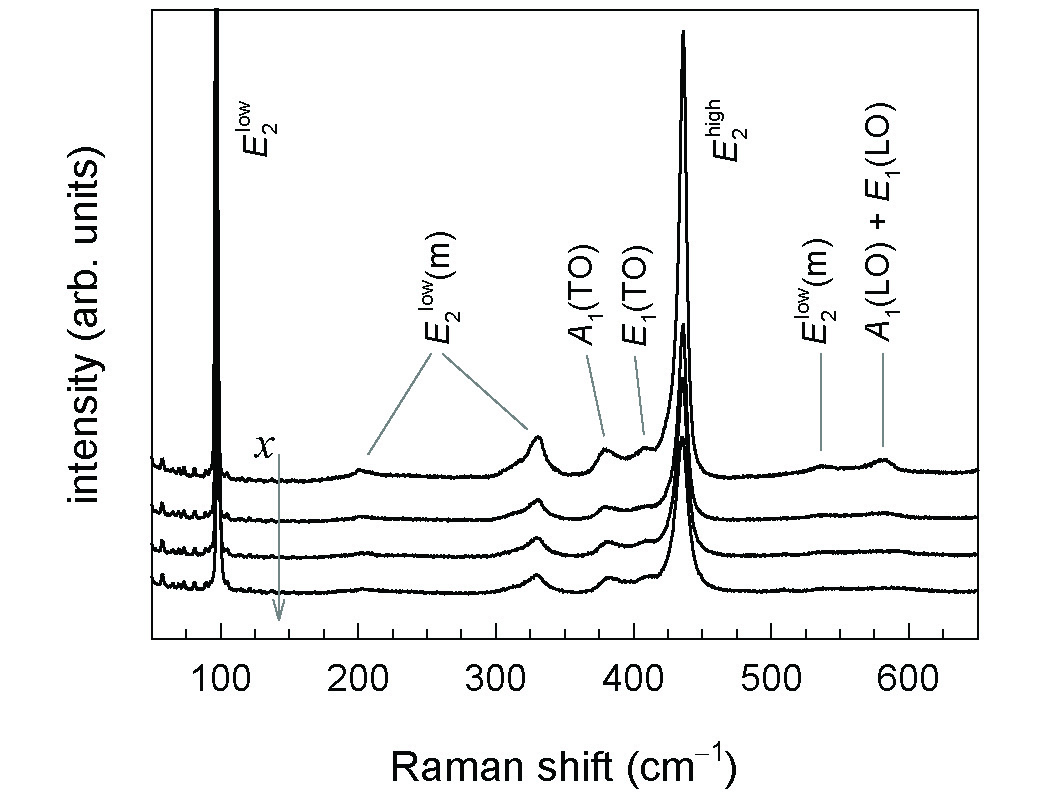} 
\caption{Raman spectra recorded at room temperature for 
\ZM\/ ($x$ = 0, 0.05, 0.10, and 0.15 from top to bottom). The excitation 
wavelength used was 514.5\,nm.} 
\label{fig:Ramanwide}
\end{figure}

Room temperature Raman spectra for \ZM\/ 
($x=0$, 0.05, 0.10, and 0.15)  are shown in Fig.\,\ref{fig:Ramanwide}, along 
with mode assignments for the observed peaks.\cite{Calleja,Damen} The 
wurtzite lattice, with space group $C_\mathrm{6v}^4$ (Hermann-Mauguin symbol 
$P6_3mc$), has four Raman-active phonon modes, $A_1+E_1+2E_2$. The two 
$E_2$ modes are nonpolar, while the $A_1$ and $E_1$ modes are polarized along 
the $z$-axis and in the $xy$-plane, respectively.\cite{Calleja,Damen,Zhang} 
The polar modes are further split into longitudinal (LO) and  transverse (TO) 
components due to the macroscopic electric field associated with the LO modes. 
Raman spectra of the ZnO and \ZM\/ compounds are dominated by two intense 
peaks of $E_2^\mathrm{low}$ ($\approx$\,98\,cm$^{-1}$) and $E_2^\mathrm{high}$
($\approx$\,436\,cm$^{-1}$) modes.  Raman scattering efficiencies
of individual modes in ZnO are known to vary with the excitation 
energy.\cite{Calleja} With 514.5\,nm (2.41\,eV) excitation, the 
highest Raman efficiencies are observed from $E_2^\mathrm{high}$ and 
$E_2^\mathrm{low}$ modes. However, the polar LO modes exhibit a strong 
resonance effect as the excitation energy approaches the electronic transition
energies. In cases when ultraviolet lasers are used for excitation, the Raman 
spectra of ZnO or \ZM\/ are dominated by the signals from LO modes.\cite{Cheng} 

\begin{figure}  \smallskip 
\centering \includegraphics[width=8cm]{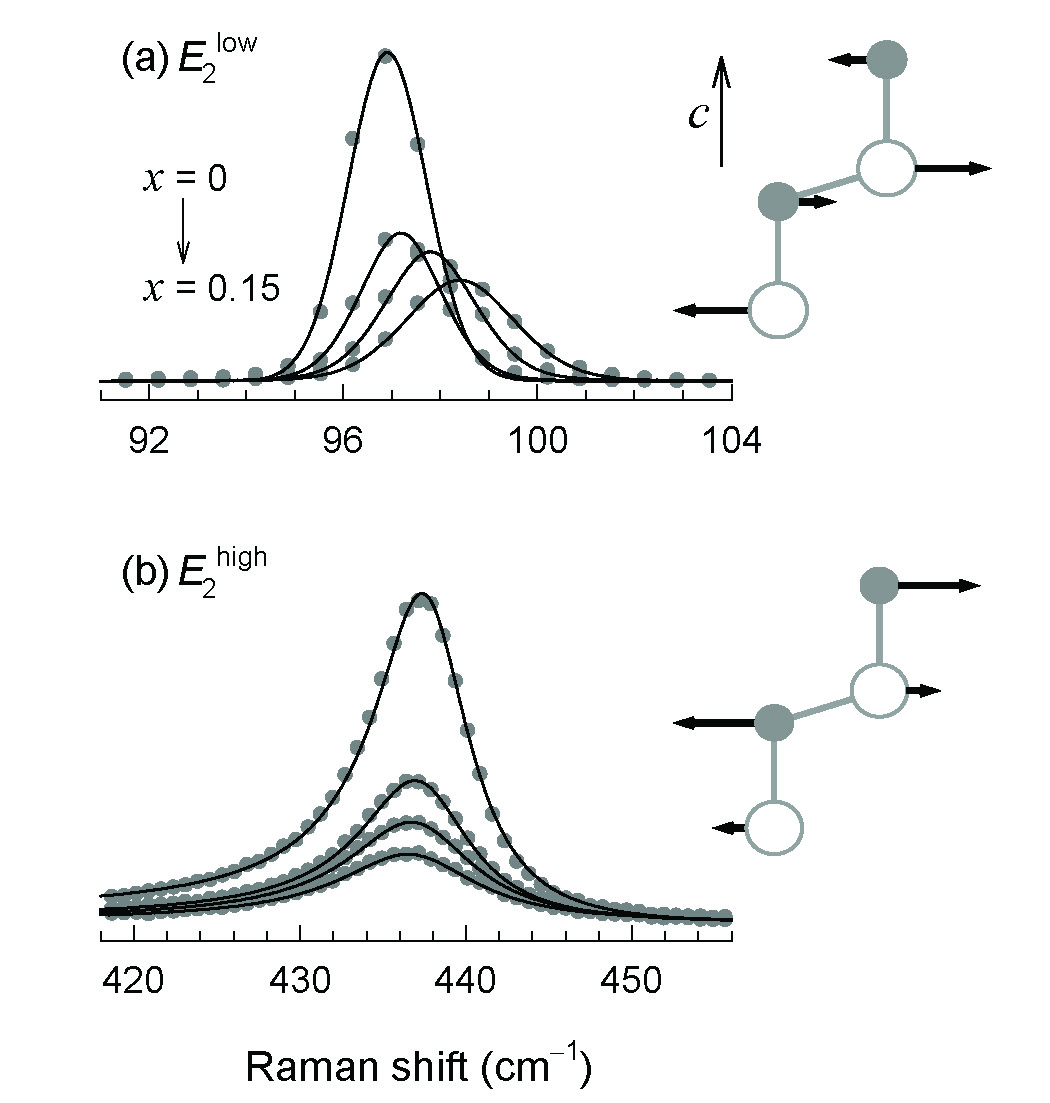} 
\caption{Line profile fittings for (a) $E_2^\mathrm{low}$ 
and (b) $E_2^\mathrm{high}$ Raman peaks of \ZM\/ (observed, gray
circles; fit, solid lines; $x$ = 0, 0.05, 0.10, and 0.15 from top to
bottom). Simplified lattice vibrations are depicted using  large open
circles for cations (Zn, Mg) and small gray circles for oxygen.} 
\label{fig:Ramanfit}
\end{figure}

\begin{figure}  \smallskip
\centering \includegraphics[width=8cm]{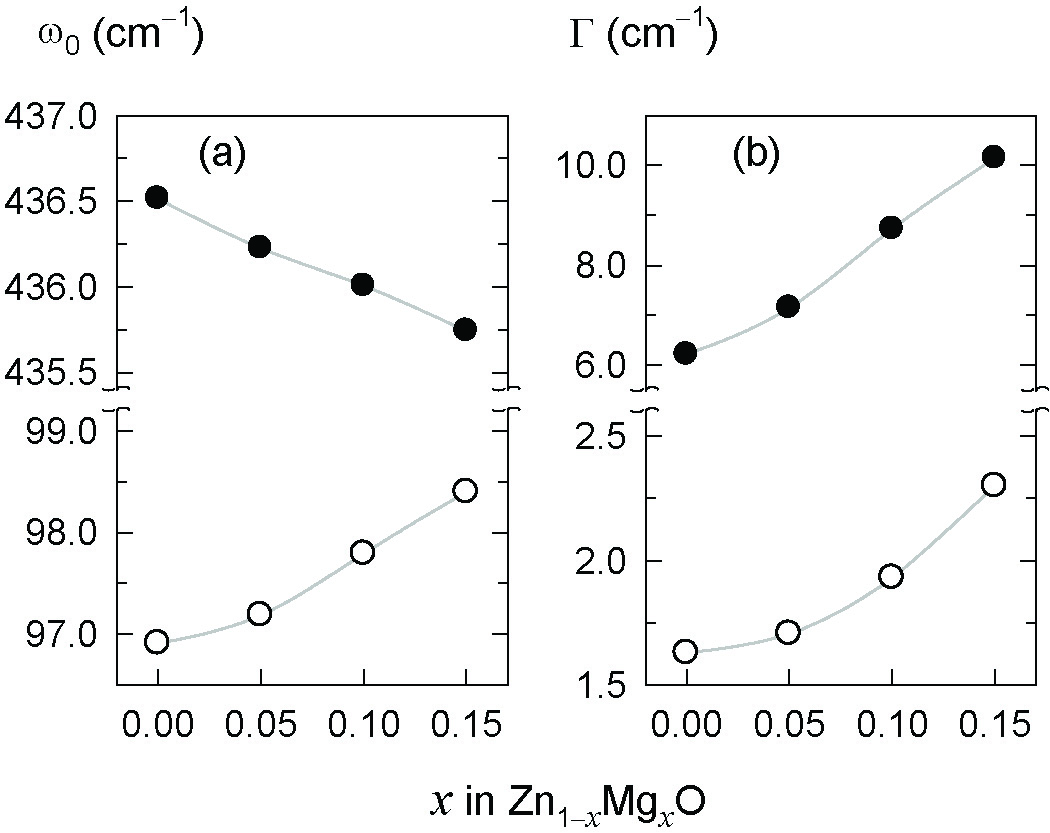} 
\caption{(a) Raman shifts $\omegaup_0$ and (b) linewidths
$\Gamma$ for  the $E_2^\mathrm{low}$ (open circles) and
$E_2^\mathrm{high}$ (filled circles)  modes of \ZM.} 
\label{fig:RamanEW}  \end{figure}

In order to quantify composition-dependent changes in the Raman spectra of \ZM, 
the $E_2^\mathrm{low}$ and $E_2^\mathrm{high}$ line profiles were 
analyzed by least-squares fitting with standard peak functions. 
From the fits, peak position ($\omegaup_0$) and 
the full-width-at-half-maximum (FWHM, $\Gamma$) linewidths were determined. 
The profile fittings of both $E_2$ modes are shown 
in Fig.\,\ref{fig:Ramanfit}, and the evolution of peak 
parameters $\omegaup_0$ and $\Gamma$ with Mg-substitution $x$ are plotted in 
Fig.\,\ref{fig:RamanEW}. Consistent with the previous study,\cite{Kim2} the 
$E_2^\mathrm{high}$ peaks  of \ZM\/ are best represented by Lorentzian 
Breit-Wigner-Fano (BWF) lineshapes with some asymmetry. 
However, the $E_2^\mathrm{low}$ line profiles do not fit as well to 
the Lorentzian-type functions. 
Instead, the $E_2^\mathrm{low}$ lines of \ZM\/ are found to be 
pseudo-Voigt-type with predominantly Gaussian components of 
97\%, 83\%, 77\%, and 81\,\% for  $x$ = 0, 0.05, 0.10, and 0.15, respectively. 

As shown in Fig.\,\ref{fig:RamanEW}a, the two $E_2$ modes of \ZM\/
exhibit  distinct dependences of phonon energy on the composition. 
With increased concentration of magnesium, the $E_2^\mathrm{low}$  mode shows
a blueshift, while the $E_2^\mathrm{high}$ mode exhibits a redshift.  For
explaining these opposing trends, the vibrational eigenvectors of the
wurtzite  $E_2$ modes need to be considered. As shown in
Fig.\,\ref{fig:Ramanfit}, both  $E_2^\mathrm{low}$ and
$E_2^\mathrm{high}$ modes are associated with atomic motions in 
the $ab$-plane. The lower energy branch corresponds  mainly 
($\approx$85\,\%) to the vibrations of heavier components  (cations,
in case of \ZM), and conversely the higher energy one corresponds 
mainly to those of lighter components (oxygen). 
Consequently, the $E_2^\mathrm{low}$ mode energy of \ZM\/ is explicitly 
affected by the cationic substitution, according to the reduced mass effect. 
Replacement of Zn with Mg will decrease the reduced mass of the
oscillator and in turn increase the phonon energy.  
By comparison, the change of cation mass should have less influence 
on the $E_2^\mathrm{high}$ mode energy, and its redshift is adequately 
attributed to the phonon softening caused by the in-plane lattice
expansion. As previously reported, the lattice constant $a$ of \ZM\/
increases monotonically with $x$,\cite{Kim1} which appears to 
account for the observed $E_2^\mathrm{high}$ mode behavior. 

Peak shapes and linewidths of Raman spectra are dictated by the underlying 
line-broadening mechanisms. 
Gaussian line broadening is intrinsic to Raman spectra and originates from 
instrumental resolution. 
Also, if the sample being examined contains inhomogeneous components with 
different phonon frequencies, the resulting Raman lineshape would have 
a Gaussian character, reflecting the statistical nature of the spectrum. 
On the other hand, Lorentzian line broadening is mostly attributed to
finite phonon lifetimes ($t$), according to the energy-time
uncertainty  relationship $\Gamma/\hbar=1/t$. The phonon
lifetime shortening, which will cause linewidth broadening, 
can occur \textit{via} two independent phonon decay mechanisms; 
by anharmonic decay into other Brillouin zone phonons or by phonon disruption 
at crystal defects. The latter, involving lattice imperfections, are
quite common for alloy and solid-solution systems. Specifically
for \ZM, it is expected that Mg-substitution will
substantially reduce the size of ordered domains, and thereby block
the long-range propagation of zone center phonons. 

As noted above, the $E_2^\mathrm{low}$ and $E_2^\mathrm{high}$ lines of
ZnO are respectively fitted with Gaussian and Lorentzian functions.
It indicates that the $E_2^\mathrm{high}$ mode of ZnO
has anharmonic  phonon decay channels, whereas the
$E_2^\mathrm{low}$ phonon does not.  The Lorentzian line broadening
behavior of $E_2^\mathrm{high}$ mode,  in Fig.\,\ref{fig:RamanEW}b,
further manifests that the Mg-substitution increasingly  populates the
phonon field of \ZM\/ with defect centers.  The $E_2^\mathrm{low}$
mode also undergoes a gradual line broadening with $x$,  and 
its peak shape changes upon Mg-substitution.  While 
unsubstituted ZnO displays a nearly pure Gaussian peak for the
$E_2^\mathrm{low}$ mode,  the peaks from \ZM\/ ($x>0$) samples are
found to contain $\approx$20\,\% of Lorentzian components.  Therefore
we conclude that, for the \ZM\/ system, the same line broadening 
mechanism applies to both $E_2^\mathrm{low}$ and $E_2^\mathrm{high}$
modes: phonon lifetime shortening by increased crystal defects
upon Mg substitution. 

\subsection{Solid-state $^{67}$Zn and $^{25}$Mg NMR spectroscopy} 
A different view of the disorder induced by Mg-substitution in 
ZnO materials can be obtained from solid-state NMR, using the NMR-active 
isotopes $^{25}$Mg (spin $I=-\frac52$) and $^{67}$Zn ($I=\frac52$). 
These nuclei are challenging to observe, because their 
low natural abundances and low gyromagnetic ratios (see Experimental) 
yield low signal sensitivities, 
and because second-order quadrupolar interactions 
associated with nuclear spins \textbar$I$\textbar\/ $>$ $\frac12$ 
can lead to a dramatic broadening of their NMR spectra. 
Second-order quadrupolar effects, however, scale with the inverse 
of the static magnetic field $B_0$, and are therefore 
mitigated at high magnetic fields, which also 
increase signal sensitivity. 

\begin{figure}  \smallskip 
\centering \includegraphics[width=8cm]{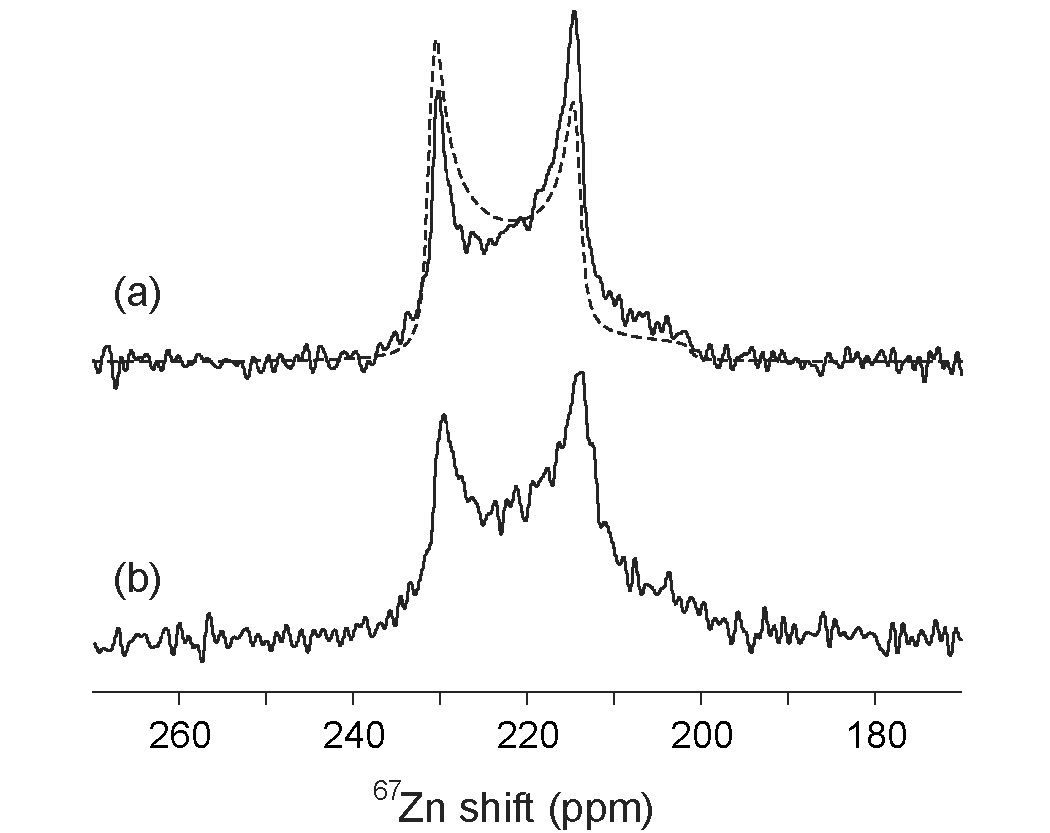} 
\caption{Solid-state spin-echo $^{67}$Zn MAS NMR spectra acquired at 19.6\,T, 
room temperature, and 10\,kHz MAS: (a) ZnO and 
(b) Zn$_{0.85}$Mg$_{0.15}$O powders, both prepared at 900$^{\circ}$C in N$_2$. 
Accompanying the ZnO spectrum in (a) is a fit (dashed line) of 
the experimental lineshape to a simulated second-order 
quadrupolar powder pattern.} 
\label{fig:nmrZn}  \end{figure}

Solid-state $^{67}$Zn MAS NMR spectra acquired at 19.6\,T of ZnO and 
Zn$_{0.85}$Mg$_{0.15}$O are shown in Fig.\,\ref{fig:nmrZn}. 
The materials were prepared by heating at 
900$^{\circ}$C in N$_2$ and yield similar spectra that clearly exhibit 
scaled second-order quadrupolar powder MAS lineshape features.\cite{Kundla} 
Associated values for the isotropic $^{67}$Zn chemical shift, $\deltaup_\mathrm{iso}$, 
the quadrupolar coupling constant, $\nuup_\mathrm{Q}$, 
and the asymmetry parameter, $\etaup_\mathrm{Q}$, can be 
determined by fitting each resolved second-order quadrupolar 
MAS lineshape. 
A simulated powder pattern is shown as the dashed line in 
Fig.\,\ref{fig:nmrZn}a accompanying the $^{67}$Zn MAS spectrum of ZnO 
and corresponding to the calculated values $\deltaup_\mathrm{iso}$ = 236\,ppm, 
$\nuup_\mathrm{Q}$ = 378\,kHz, and $\etaup_\mathrm{Q}$ = 0.1. 
Several discrepancies between the calculated and 
experimental powder patterns may be due to modest anisotropy of the powder particles 
and/or to uneven excitation of the different crystallite orientations 
by the radiofrequency pulses. 
The powder pattern and fit parameters are nevertheless consistent with 
a single crystallographically distinct and relatively symmetric 
$^{67}$Zn site in the wurtzite ZnO structure. 
The $^{67}$Zn MAS spectrum of Zn$_{0.85}$Mg$_{0.15}$O in Fig. 4b 
exhibits a similar lineshape and features, which are broadened by the 
presumably random incorporation of Mg atoms within the Zn-rich 
wurtzite lattice.

The close similarity between the $^{67}$Zn MAS NMR spectra of ZnO 
(Fig.\,\ref{fig:nmrZn}a) and Zn$_{0.85}$Mg$_{0.15}$O 
(Fig.\,\ref{fig:nmrZn}b) indicates that the substitution of 
15\,\% of Zn atoms by Mg has little influence on the local 
electronic environments of a majority of the $^{67}$Zn nuclei. 
Assuming that Mg atoms are distributed randomly, the probability 
of having no Mg atom in the second coordination sphere of a given 
$^{67}$Zn nucleus is only 14\,\% (the first coordination sphere 
consisting of four O atoms). Significant effects on the 
$^{67}$Zn spectrum would be expected if the quadrupolar interactions 
were substantially perturbed by the substitution of Mg atoms for 
1, 2, 3, 4 (probabilities of 30\,\%, 29\,\%, 17\,\%, and 7\,\%, respectively) 
or more of the 12 Zn atoms in the second coordination sphere. 
However, despite the high probability of modifications of the local 
chemical environment of the $^{67}$Zn nuclei upon 15\,\% Mg-substitution, 
dramatic changes of the $^{67}$Zn spectrum are not observed, rather only 
modest broadening that reflects increased disorder. 
Under these conditions, the 
differences induced in ZnO or Zn$_{0.85}$Mg$_{0.15}$O by changing the 
synthesis atmosphere (\textit{e.g.}, air, O$_2$, or N$_2$) and temperature 
(over the range 550$^{\circ}$C to 900$^{\circ}$C) 
were too subtle to be clearly reflected in the corresponding 
$^{67}$Zn MAS NMR spectra (data not shown), which are similar to those in 
Fig.\,\ref{fig:nmrZn}a, b. 
Thus, it appears that $^{67}$Zn NMR measurements of these ZnO-related materials 
are sensitive principally to substitutions that result in modifications 
of the first coordination sphere of the zinc nuclei (substitution of the anionic species), or to 
substitutions resulting in larger lattice distortions or increased electronic 
disorder than observed here.\cite{Toberer} 
 
\begin{figure} \smallskip
\centering \includegraphics[width=8cm]{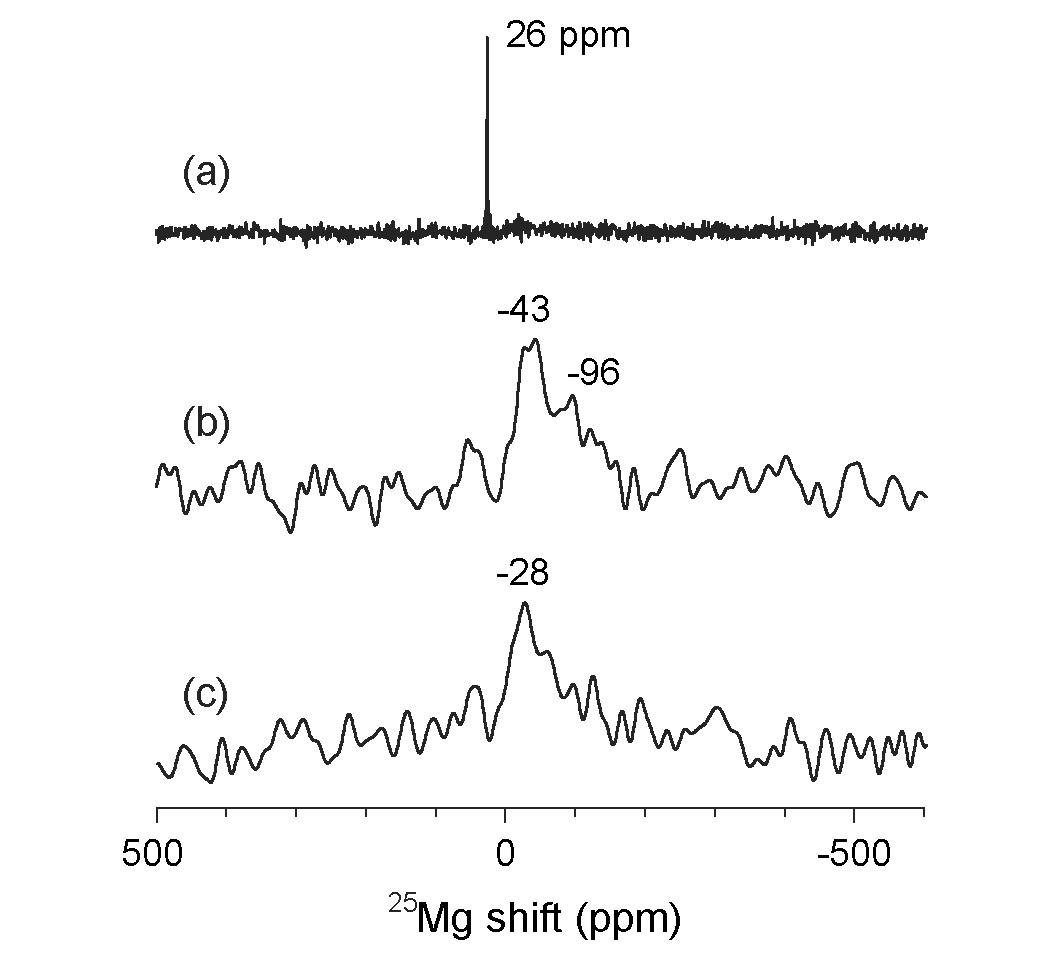} 
\caption{Spin-echo $^{25}$Mg MAS NMR spectra, recorded at 19.6\,T, of 
(a) polycrystalline MgO powder (external reference for $^{25}$Mg NMR shift), 
(b) Zn$_{0.85}$Mg$_{0.15}$O powder prepared at 550$^{\circ}$C in air, and 
(c) Zn$_{0.85}$Mg$_{0.15}$O powder prepared at 900$^{\circ}$C in N$_2$.} 
\label{fig:nmrMg} 
\end{figure}

Solid-state $^{25}$Mg MAS NMR measurements complement the $^{67}$Zn NMR 
results by being sensitive to the structural and electronic perturbations 
experienced directly by Mg atom incorporated into ZnO lattices. 
The $^{25}$Mg MAS NMR measurements were similarly conducted 
at 19.6\,T and the resulting spectra are 
shown in Fig.\,\ref{fig:nmrMg} for polycrystalline MgO and 
different Zn$_{0.85}$Mg$_{0.15}$O powders prepared at 550$^{\circ}$C in air and at 
900$^{\circ}$C in N$_2$. 
The $^{25}$Mg MAS spectrum in Fig.\,\ref{fig:nmrMg}a of MgO 
shows a narrow well-defined $^{25}$Mg peak that reflects the highly symmetric 
coordination environment of the single type of Mg site in its 
rock-salt structure.\cite{MacKenzie1} 
By comparison, the $^{25}$Mg MAS spectra of the different 
Zn$_{0.85}$Mg$_{0.15}$O powders contain broad and relatively unstructured 
lineshapes centered at \textit{ca.} $-$30\,ppm to $-$40\,ppm 
that reflect broad distributions of signal intensity and 
thus broad distributions of local $^{25}$Mg environments. 
The first Mg coordination sphere (MgO$_4$) 
in Zn$_{0.85}$Mg$_{0.15}$O is expected to be significantly distorted from 
its regular tetrahedral geometry, and the second coordination sphere, 
composed of Mg(Zn,Mg)$_{12}$, is expected to have an even larger number of 
different local configurations, bond distances, and/or bond angles. 
The $^{25}$Mg MAS NMR spectra appear to be more sensitive than the 
$^{67}$Zn NMR results to local material environments in 
Zn$_{0.85}$Mg$_{0.15}$O.  
However, the low signal-to-noise and broad lines observed for natural 
abundance $^{25}$Mg in Zn$_{0.85}$Mg$_{0.15}$O, even at 19.6\,T, 
preclude a detailed and reliable analysis of the isotropic shift distributions 
and second-order quadrupolar broadening that would be necessary to extract 
directly quantitative information on nature of the Mg site distributions 
and/or their disorder. 
It is noteworthy that the Zn$_{0.85}$Mg$_{0.15}$O powders do not 
display any spectral feature corresponding to bulk MgO, thereby ensuring the phase 
purity of \ZM\/ ($x\,\leqslant\,0.15$) solid solutions, within the 
sensitivity limit of the $^{25}$Mg NMR measurements.\cite{MacKenzie2} 

In common for $^{29}$Si, $^{27}$Al, and $^{25}$Mg, an increase of the 
coordination number for a given cationic center counts 
for enhanced local shielding, resulting in an upfield displacement 
of the signal(s) to lower shift values.\cite{Dupree,Magi} 
The spinel MgAl$_2$O$_4$ containing MgO$_4$ fragments exhibits 
a $^{25}$Mg shift of 52\,ppm, compared to 26\,ppm for MgO which is 
comprised of MgO$_6$ local units.\cite{Dupree} 
By comparison, clay minerals in which Mg atoms are coordinated by six (O,OH) 
ligands yield $^{25}$Mg shifts in the range 
of 0 to $-$100\,ppm.\cite{MacKenzie1} The wurtzite oxide 
Zn$_{0.85}$Mg$_{0.15}$O has a 4-coordinated geometry of Mg, similar to 
MgAl$_2$O$_4$, but the former clearly shows a more upfield-shifted 
$^{25}$Mg resonance at $-30\sim-$40\,ppm. 
We infer that $^{25}$Mg MAS NMR signals are considerably 
influenced by the ligand identity beyond the first coordination shell, 
and that Zn has a far stronger shielding contribution than Al, Si, 
or alkaline-earths, thus accounting for the observed resonance values. 

\subsection{Neutron diffraction}

In previous studies, the crystal structure of ZnO has been reported
[space group $P6_3mc$, Zn at $(\frac13\frac230)$ and O at 
$(\frac13\frac23u)$] with lattice constants of 
$a=$ 3.2427$\sim$3.2501\,\AA\/ and $c=$ 5.1948$\sim$5.2071\,\AA, and 
atomic position parameter  
$u=$ 0.381$\sim$0.3826.\cite{Sawada,Sabine,Abrahams,Harrison,Albertsson,Kisi} 
In our previous synchrotron x-ray study  on polycrystalline ZnO, we obtained 
$a=$ 3.2503\,\AA, $c=$ 5.2072\,\AA, and $u=$ 0.3829.\cite{Kim1} 
Here, we reexamine the crystal structures of polycrystalline ZnO and 
Zn$_{0.875}$Mg$_{0.125}$O utilizing neutron scattering, which has several 
advantages over x-rays.  The coherent neutron scattering length of 
O (5.804\,fm) is comparable to those of Zn (5.680\,fm) and 
Mg (5.375\,fm),\cite{Sears} as distinct from x-ray diffraction, where 
scattering is strongly weighted by Zn. 
Moreover, neutron scattering can provide 
high $Q$ diffraction data with much less attenuation than x-rays, due to there 
being no fall-off with $Q$ in the form factor.

\begin{figure}  \smallskip 
\centering \includegraphics[width=8cm]{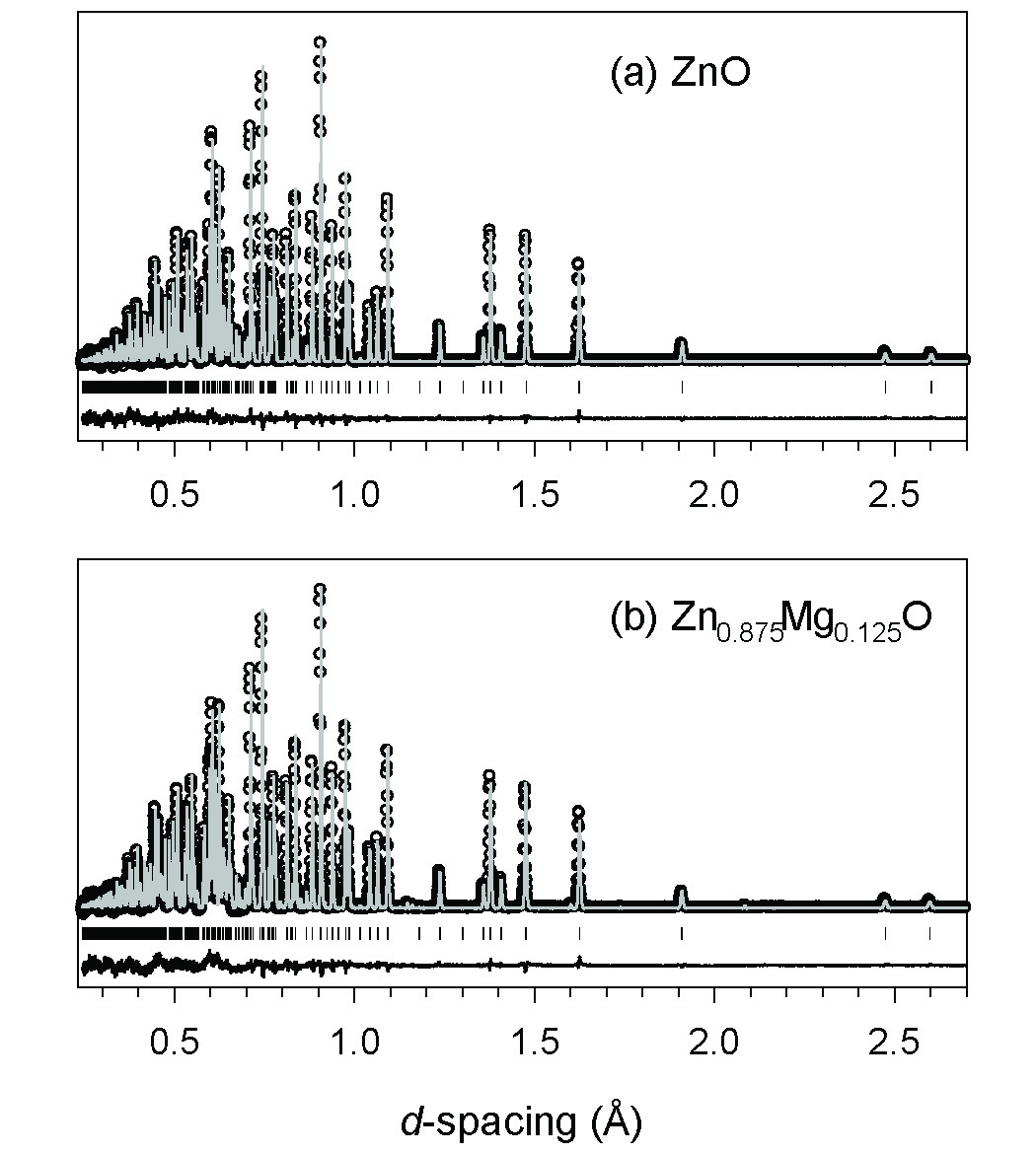} 
\caption{Rietveld refinement of time-of-flight powder neutron diffraction 
profiles of (a) ZnO and (b) \ZMM. Calculated patterns (gray lines) are 
superimposed on observed data (open circles), with Bragg positions and 
difference profiles at the bottom.}  \label{fig:Riet} 
\end{figure}

\begin{table}[b] \caption{Structural parameters for ZnO and
Zn$_{0.875}$Mg$_{0.125}$O  determined from the Rietveld refinement of
time-of-flight  neutron diffraction pattern.} \begin{ruledtabular}
\begin{tabular}{lll} & ZnO  &Zn$_{0.875}$Mg$_{0.125}$O  \\ \hline $V$
(\AA$^3$)	 & 47.603(1)  & 47.578(1) \\ $a$ (\AA)      &
3.24945(1) & 3.25058(2) \\ $c$ (\AA)      & 5.20574(3) & 5.19771(3) \\
$c/a$          & 1.60204    & 1.59940    \\ $u$            &
0.38275(5) & 0.38214(6) \\ $U_\mathrm{iso}$(Zn/Mg) (\AA$^2$) &
0.00648(8) & 0.0065(2) \\ $U_\mathrm{iso}$(O) (\AA$^2$) & 0.0101(1)  &
0.0087(2) \\ $d_{M-\mathrm{O}}$ (\AA) & (1$\times$) 1.9925(3) &
(1$\times$) 1.9868(3) \\ & (3$\times$) 1.9729(1) & (3$\times$)
1.9743(1) \\ $\angle_{\mathrm{O}-M-\mathrm{O}}$ ($^\circ$) &
108.022(7) & 108.081(9) \\ & 110.881(7)	& 110.825(9) \\ \end{tabular}
\end{ruledtabular} \label{tab:Riet} \end{table} 

\begin{figure}  \smallskip 
\centering \includegraphics[width=8cm]{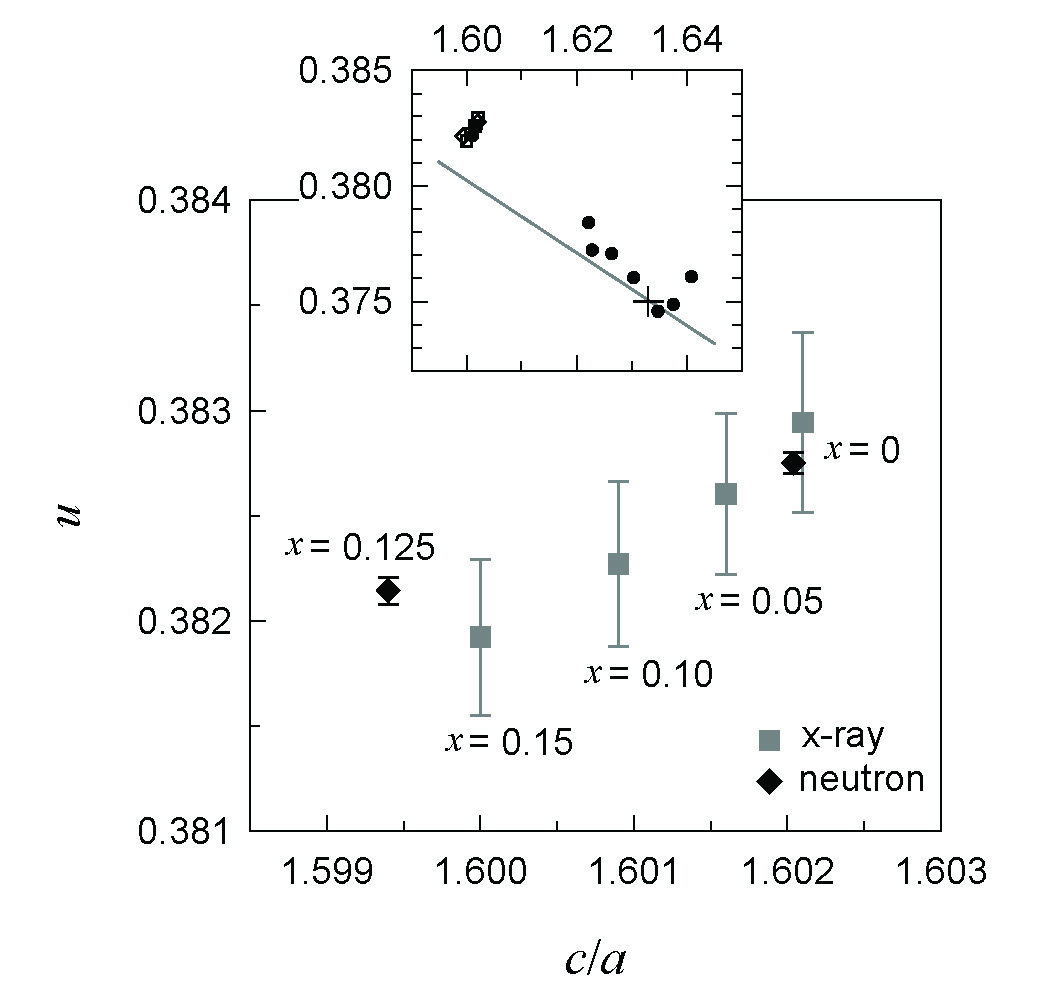} 
\caption{Average wurtzite parameters $c/a$ and $u$ of ZnO and
\ZM\/ determined by Rietveld refinements of neutron (diamonds) and
synchrotron x-ray (squares, ref.\,\cite{Kim1}) diffraction patterns. 
Note the significantly smaller error bars that emerge from the neutron study. 
The inset shows the data observed from existing wurtzites (see
ref.\,\cite{Kim1} for details).}  \label{fig:ca_u}  \end{figure}

Rietveld refinements of ZnO and Zn$_{0.875}$Mg$_{0.125}$O were carried out on
four histograms of data collected at detector locations of 46$^{\circ}$, 
90$^{\circ}$, 119$^{\circ}$, and 148$^{\circ}$. The highest-$Q$ data
from the 148$^{\circ}$ detector covers the range of $d>0.25\,\AA$ 
($Q<25\,\AA^{-1}$). The model used for the refinement in space group $P6_3mc$ 
has Zn/Mg at the $2b$ Wyckoff position ($\frac13\frac230$) and O also at 
2$b$ position ($\frac13\frac23u$). For Zn$_{0.875}$Mg$_{0.125}$O, Zn and Mg 
were statistically distributed over the common site with fractional 
occupancies fixed to the respective compositions. 
The refinement converged with reliability factors of 
$R_\mathrm{wp}=2.57\,\%, R_\mathrm{p}=1.80\,\%$, and $\chiup^2=2.71$ 
for ZnO, and $R_\mathrm{wp}=2.89\,\%, R_\mathrm{p}=2.04\,\%$, and 
$\chiup^2=3.56$ for Zn$_{0.875}$Mg$_{0.125}$O. 
The Rietveld refinement profiles and the structural parameters are given 
in Fig.\,\ref{fig:Riet} and Table\,\ref{tab:Riet}, respectively. 
The present refinement results agree well with the findings from our 
previous synchrotron x-ray study.\cite{Kim1} 
Both studies indicate that Mg-substitution expands the $a$-parameter, 
compresses the $c$-parameter, and decreases the oxygen position 
parameter $u$, as summed up in Fig.\,\ref{fig:ca_u}. 
The two data sets from x-ray and neutron reveal an identical trend in the
structural evolution in terms of $a$, $c$, and $u$. 

\begin{figure}\smallskip\centering \includegraphics[width=8cm]{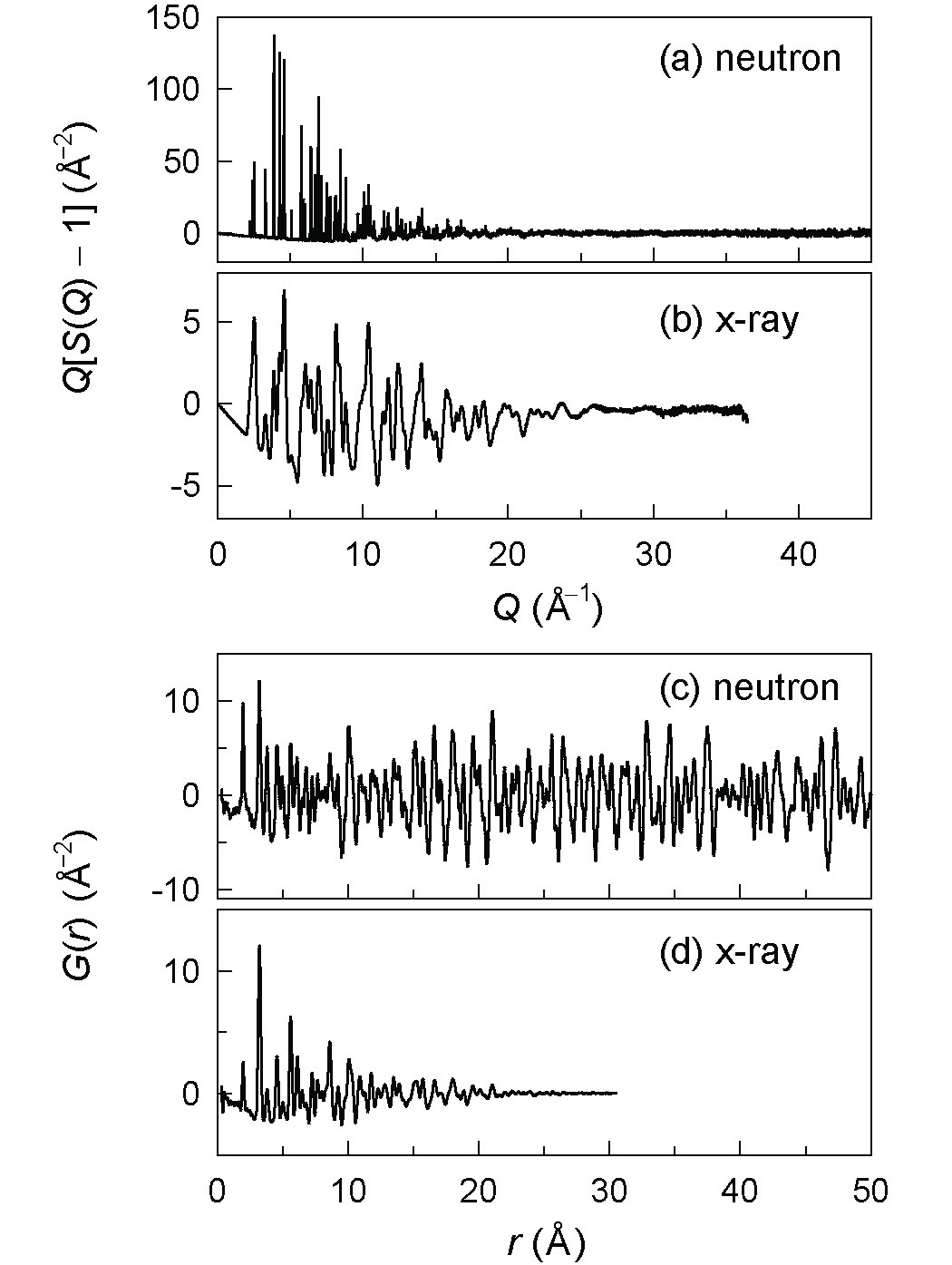} 
\caption{Comparison of high-$Q$ scattering data from ZnO measured 
using synchrotron x-ray and pulsed-neutron radiation: 
(a,b) $F(Q)=Q[S(Q)-1]$, and (c,d) $G(r)$. 
In the Fourier transform of $F(Q)$ to $G(r)$, $Q$ data were terminated 
at 28\,\AA$^{-1}$ for x-rays, and 35\,\AA$^{-1}$ for neutrons.} 
\label{fig:SFG}  \end{figure}

For the composition range up to $x=0.15$, the \ZM\/ solid solutions remain 
isostructural with wurtzite ZnO, with both Zn and Mg tetrahedrally 
coordinated by O atoms. However, since the two cations Zn and Mg have 
clearly distinct crystal chemistries in oxidic environments, it is
conjectured  that they may have distinct local geometries within 
\ZM\/ lattices.  In order to directly address this question, we
have carried out neutron PDF analyses of 
Zn$_{0.875}$Mg$_{0.125}$O and ZnO. Figure\,\ref{fig:SFG} shows the
PDF data for ZnO, comparing the present neutron data with the x-ray
data previously measured at Beamline 11-ID-B of the Advanced Photon Source 
at Argonne National Laboratory. As can be seen from Figs.\,\ref{fig:SFG}a 
and \ref{fig:SFG}b, the neutron study provides scattering information 
over a significantly higher $Q$ range and also with higher resolution. 
Correspondingly, the neutron PDF
$G(r)$ can be obtained for far wider $r$-ranges than when using 
x-rays (Fig.\,\ref{fig:SFG}c). Given this, neutrons are 
expected to enable better description of the \ZM\/ alloy structure(s).

\begin{figure} 
\centering \includegraphics[width=8cm]{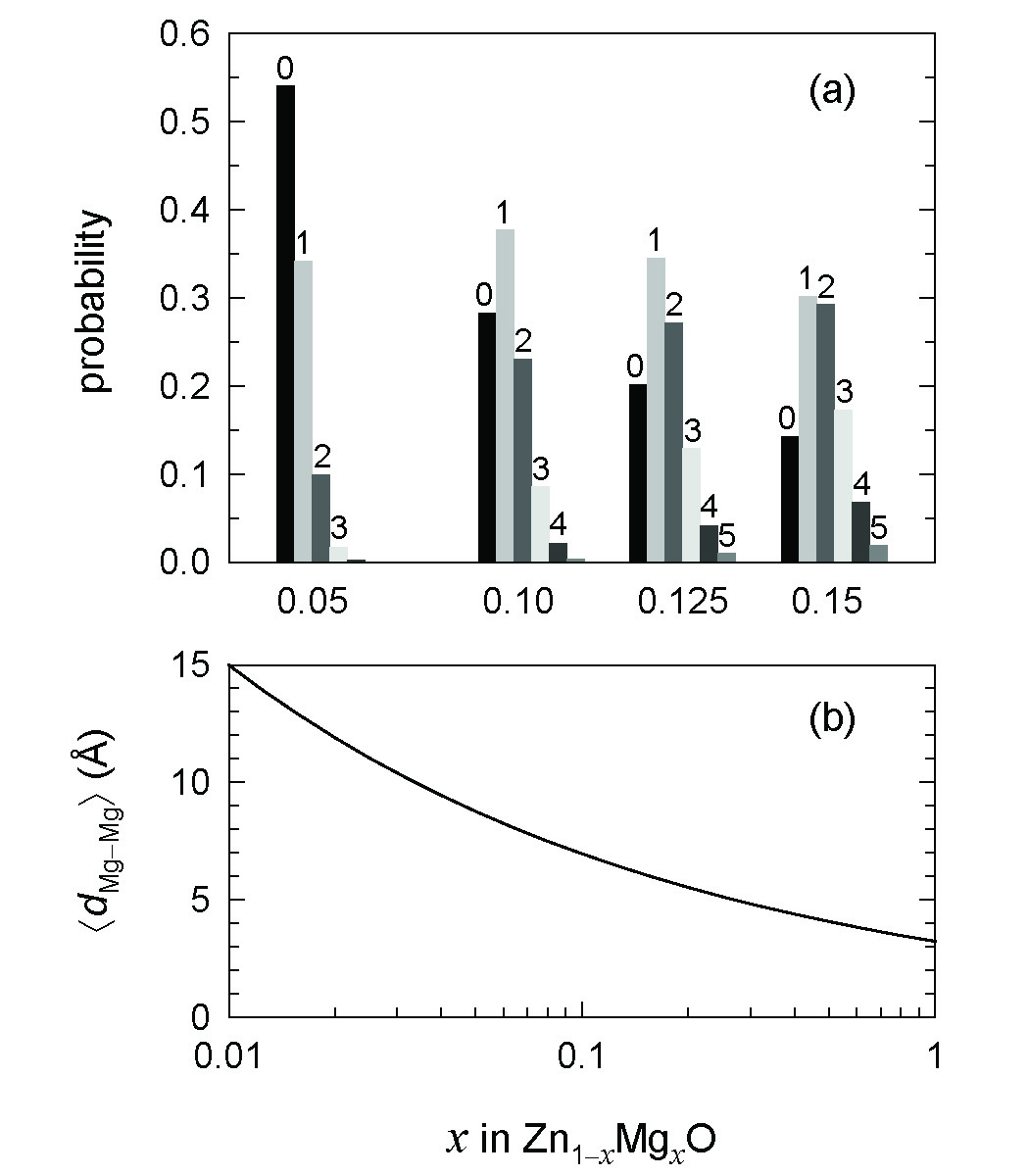} 
\caption{(a) Probability for a cation in \ZM\/ to have 0,
1, 2, 3, 4  or 5 Mg neighbors in the 12-membered coordination
shell ($r\approx$ 3.25\,\AA), and (b) mean separation between the
nearest  Mg$-$Mg pairs.}  \label{fig:Mgdist}  \end{figure}

In order to obtain an appropriate real-space model for the PDF analysis
of Zn$_{0.875}$Mg$_{0.125}$O,  we estimate as follows, 
the impact of Mg-substitution on the chemical environment 
within the ZnO lattice. 
Cations in the wurtzite lattice form a hexagonal 
close-packed sublattice, where each cation has 12 nearest neighbors. 
At the Mg-substitution level of $x=$ 0.125, it is estimated 
that $\approx$80\,\% of the cations have at least one Mg 
neighbor in the cation sublattice (Fig.\,\ref{fig:Mgdist}).  
Each cation (whether Zn or Mg) has as neighbors, 1.5 Mg atoms 
on average, and the mean Mg$-$Mg distance is as short 
as 6.5\,\AA. Therefore the coordination geometry of MgO$_4$ will 
serve a substantial structural factor in the PDF of \ZMM\/ and 
needs to be treated with independent parameters in the PDF refinement.  

\begin{figure}  \smallskip \centering 
\centering \includegraphics[width=8cm]{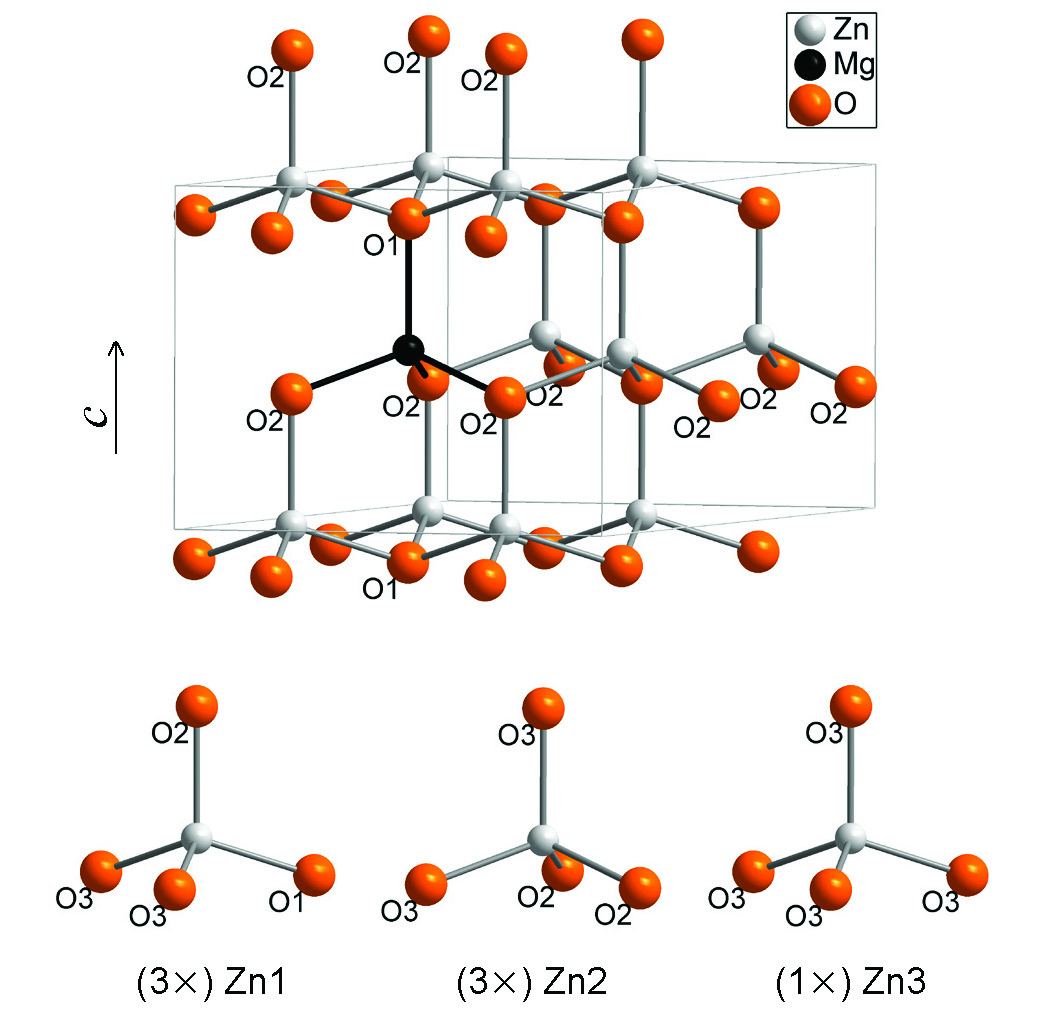} 
\caption{(Color online) Zn$_7$MgO$_8$ supercell model
structure used for neutron PDF analyses of \ZMM,
and  coordination environments for three distinct Zn sites therein. 
In the top part, unlabeled oxygen atoms belong to the O3 type (see
text).}  \label{fig:supercell}  \end{figure}

On the basis of the above considerations, the structural model for
\ZMM\/ was designed as follows. To distinguish the 
Mg and Zn atoms in the unit cell, while maintaining the composition, a
2\,$\times$\,2\,$\times$\,1 supercell of the primitive wurtzite structure 
was selected, with the cell parameters $a=b\approx$ 6.5\,\AA, 
$c\approx$ 5.205\,\AA, $\alphaup=\betaup=$ 90$^{\circ}$, 
$\gammaup=$ 120$^{\circ}$,  and the unit content Zn$_7$MgO$_8$. 
In the average structure scheme, in which the atomic positions are set as 
($\frac13\frac230$) for cations and ($\frac13\frac23u$) for  anions, 
the external and internal geometries of all the tetrahedra are uniformly 
defined by the $c/a$ ratio and $u$, respectively. 
However in the supercell model, we allowed the MgO$_4$ unit to have 
its own coordination  geometry, irrespective of the ZnO$_4$ geometry. 
As a result, the eight O atoms in the supercell unit are divided into 
three groups; one O atom bonded to Mg along the $c$-axis (O1), 
three O atoms bonded to Mg laterally (O2), 
and the remaining four O atoms (O3). The seven Zn
atoms are also grouped according to their proximities to 
the Mg atom; three Zn atoms having an apical O2 oxygen atom (Zn1), 
three Zn atoms bonded to O2 atoms laterally (Zn2), and one Zn bonded to 
only O3 atoms (Zn3). 
Figure\,\ref{fig:supercell} illustrates the supercell configuration
and the  three different types of ZnO$_4$ geometries. 

The PDF of ZnO was analyzed using a simple wurtzite cell
(Zn$_2$O$_2$) obeying the symmetry requirements of the $P6_3mc$
space group.  However, for the supercell refinement of \ZMM, 
several symmetry constraints were lifted from the average structure 
description. 
The $z$-coordinates of the O1, O2, and O3 types of oxygen atoms were 
independently varied, and the $z$-coordinate of Mg was also refined. 
We also attempted in-plane displacements of the O2 atoms, but found 
that these were unstable in the refinement. 
Therefore, the refinement for \ZMM\/ included three more position 
parameters than those for ZnO. 
For both ZnO and \ZMM, the refinements of $G(r)$ used the variables of 
lattice constants, atomic position parameters, isotropic temperature factors,  
structure scale factor, and peak sharpening coefficients. 

\begin{figure}  \smallskip \centering 
\centering \includegraphics[width=8cm]{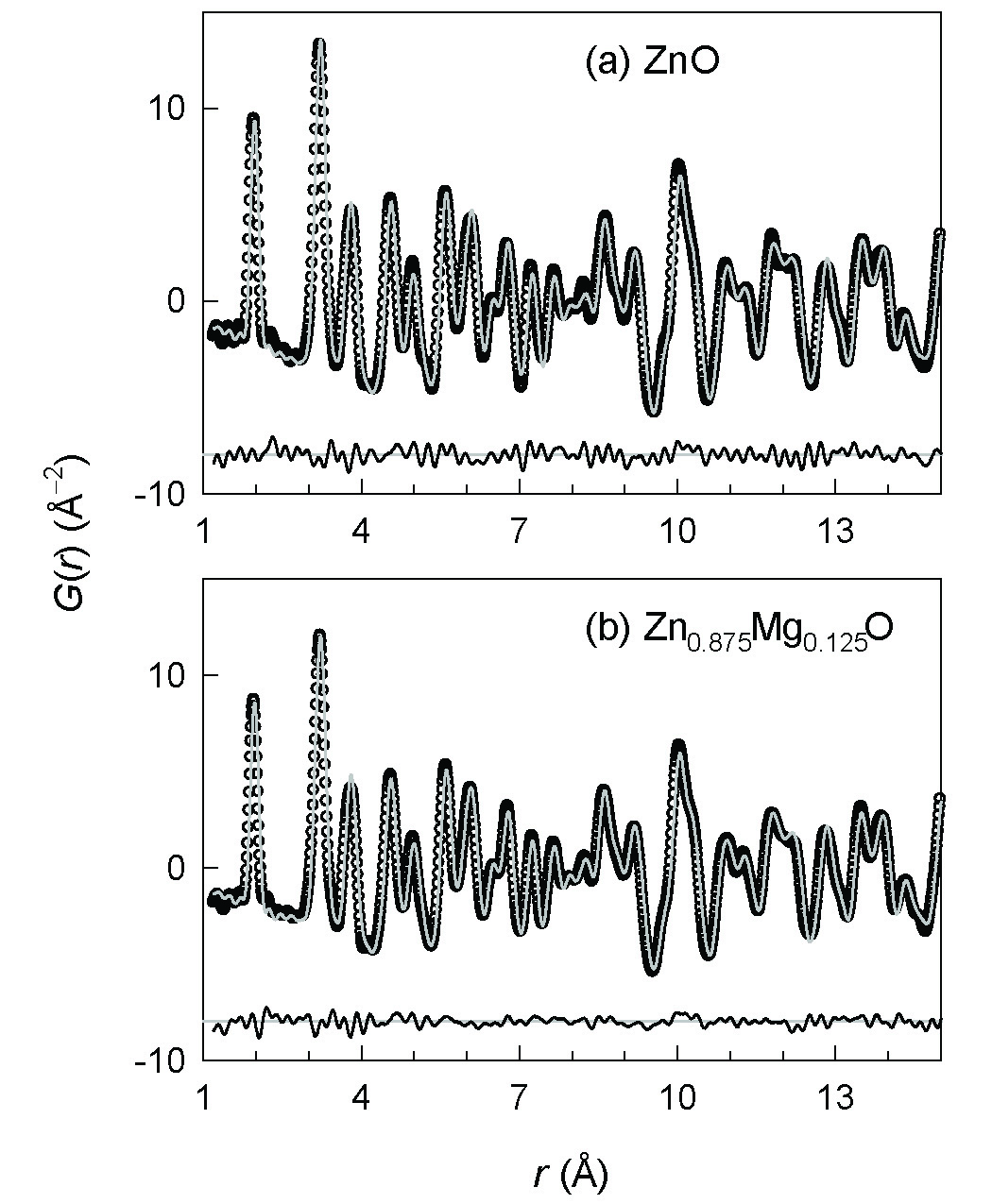} 
\caption{Neutron PDF refinements for (a) ZnO and (b) \ZMM, 
parts of the 1.2$-$22\,\AA\/ fittings. 
Calculated PDFs (gray lines) are superimposed on experimental data
(open circles), with the differences shown at the bottom of each.}  
\label{fig:pdffit} \end{figure}

Figure\,\ref{fig:pdffit} shows the neutron PDF refinements for ZnO
and \ZMM, carried out in the $r$ range of 1.2\,\AA\/ to 22\,\AA.  
The refinement parameters are summarized in Table\,\ref{tab:PDF}. 
For both structures, the refinements were achieved with satisfactorily 
low $R_\mathrm{w}$'s  of 7.9\,\% (ZnO) and 8.3\,\% (Zn$_{0.875}$Mg$_{0.125}$O). 
In comparison with the Rietveld results, the PDF-refined lattice constants 
are systematically larger by $\approx$0.06\,\%, but good agreements are 
observed for the $c/a$ ratios. Also, the oxygen atom position 
in ZnO is reasonably well reproduced from the PDF and 
Rietveld refinements. 
An interesting finding emerges from the atomic coordinates of Mg and 
its adjacent oxygen atoms (O1 and O2) in \ZMM. 
As found in Table\,\ref{tab:PDF}, the three groups of O atoms have 
clearly distinct $z$-coordinates from one another. 
Both the O1 and O2 groups are vertically shifted towards their respective 
Mg atoms. 
Mg itself also moves towards the basal plane formed of O2 atoms. 
The resulting MgO$_4$ unit has a squashed tetrahedral geometry (point 
group $C_\mathrm{3v}$), with a much shorter height than those of 
ZnO$_4$ tetrahedra. However the O3 atoms, which are not directly bonded to 
Mg, have similar $z$-coordinates to those found in ZnO. Therefore, the 
(Zn3)(O3)$_4$ tetrahedra retain the unperturbed geometry of ZnO. 
The other types of ZnO$_4$ tetrahedra share corners with MgO$_4$ and 
are likely to undergo intermediate distortion. 

\begin{table}[b] \caption{Structural parameters from the neutron PDF
refinements of  ZnO and \ZMM, over the $r$ range
of 1.2$-$22\,\AA.} 
\begin{ruledtabular} 
\begin{tabular}{lll} 
& ZnO       &Zn$_{0.875}$Mg$_{0.125}$O  \\ 
\hline
$a$ (\AA) & 3.2514(3) & 3.2527(4) $\times$\,2 \\ 
$c$ (\AA) & 5.2089(6) & 5.2015(8)          \\
$c/a$     & 1.6020    & 1.5991             \\  
$z$(O)    & 0.3824(4)/0.8824(4) & 0.3825(8)/0.8825(8) (O3) \\ 
          &           & 0.386(3) (O2)      \\
          &           & 0.858(5) (O1)      \\  
$z$(Mg)   &           & 0.486(8)           \\  
$U_\mathrm{iso}$(Zn/Mg) (\AA$^2$) & 0.0086(8) & 0.0063(6) \\ 
$U_\mathrm{iso}$(O) (\AA$^2$) & 0.0070(6) & 0.0081(8) \\ 
\end{tabular} 
\end{ruledtabular} 
\label{tab:PDF}
\end{table} 

\begin{figure}  \smallskip 
\centering \includegraphics[width=8cm]{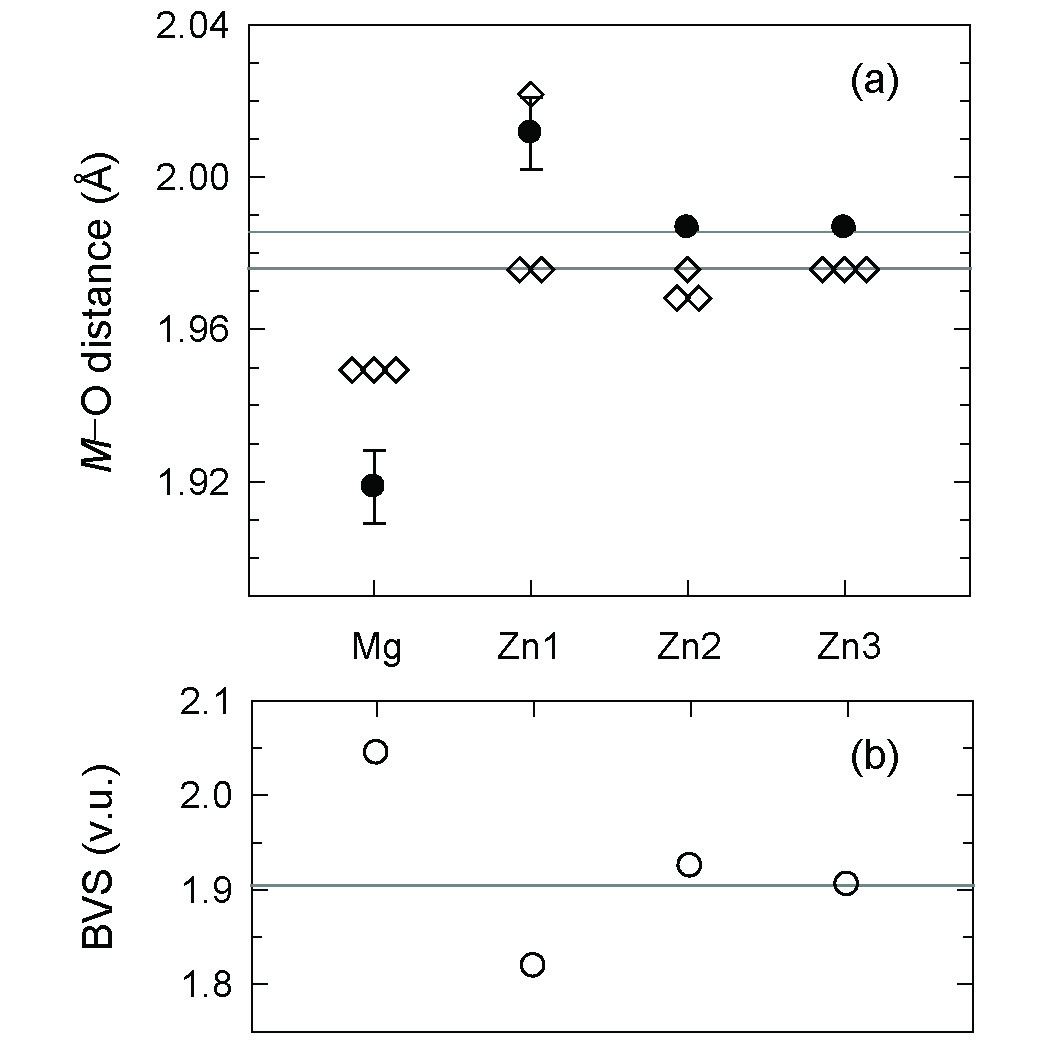} 
\caption{(a) Mg$-$O and Zn$-$O bond distances in \ZMM, 
as analyzed by the neutron PDF refinement of a Zn$_7$MgO$_8$ 
supercell  model ($c$-axial bonds, circles; lateral 
bonds, diamonds), and (b) bond valence sums for each cation site. 
Horizontal lines indicate the Zn$-$O bond distances (in the upper
panel) and BVS for Zn (in the lower panel) in ZnO.} 
\label{fig:bondl}  \end{figure}

To examine the presence of short-range structural characteristics, 
the PDF refinements were carried out from 1.2\,\AA\/ to different 
values of $r_\mathrm{max}$ between 10$-$30\,\AA. 
The lattice constants and position parameters showed insignificant 
variations with $r_\mathrm{max}$, and the $R_\mathrm{w}$ values 
had shallow minima at $r_\mathrm{max}\approx22\,\AA$ for both ZnO 
and \ZMM. 
The Mg$-$O and Zn$-$O bond distances plotted in Fig.\,\ref{fig:bondl} 
represent the averaged results of multiple refinements over
1.2$-r_\mathrm{max}$\,(\AA), where $r_\mathrm{max}$ was increased from
10 to 30\,\AA\/ in 2\,\AA\/ steps. 
From Fig.\,\ref{fig:bondl}a, we find a meaningful distinction between 
the MgO$_4$ and ZnO$_4$ bonding structures. 
The Mg$-$O bonds are shorter than the Zn$-$O bonds, and the
former set consists of one short and three longer bonds, in contrast
to the latter. 
These results are consistent with the greater sensitivity of 
the $^{25}$Mg NMR spectra (Fig.\,\ref{fig:nmrMg}b, c) to 
Mg substitution into ZnO-rich lattices, compared to the  
$^{67}$Zn NMR results (Fig.\,\ref{fig:nmrZn}).
For Mg and the three Zn atom types in \ZMM, 
bond valence sums (BVS)\cite{Brown1} are evaluated, as shown 
in Fig.\,\ref{fig:bondl}b.  
The valence sums for all of the cation and anion sites in \ZMM\/ are
within a reasonable range around the ideal magnitude of 2 valence
units (v.u.). 
The relatively underbonded situation for Zn1 atoms and the 
slight overbonding of the Mg atom are the results of a rather  
drastic shift of the bridging O1 atom. 
The global instability index (GII)\cite{Brown1} 
of \ZMM\/ is calculated as 0.116\,v.u., which is
significantly larger than that of ZnO (0.067\,v.u.). 
The GII value is often used as a measure of the residual bond strain, 
and is known not to exceed 0.2\,v.u. for ordinary structures 
in the standard state.\cite{Brown1,Brown2}  
It is worth mentioning, in light of the above GII
considerations, that the increased microstrain upon Mg-substitution 
results in characteristic optical and x-ray line broadening 
behavior of \ZM\/ that we have previously reported.\cite{Kim3,Kim4} 

The solid solution of \ZM\/ has the thermodynamic solubility limit of 
$x\approx$ 15\,\% on the ZnO-rich side. Obviously the distinct
coordination preferences of Mg and Zn prevent the formation of 
a continuous solid solution across the entire composition range, $x$. 
While Mg has a point symmetry of $O_\mathrm{h}$ in its binary oxide MgO, 
wurtzite ZnO possesses only a $C_\mathrm{3v}$ environment. 
Here, we rationalize the neutron PDF result for the MgO$_4$ tetrahedral 
geometry, in several ways. 
Both bond valence\cite{Brese} and ionic radii\cite{Shannon} compilations 
are useful to explain the smaller tetrahedral volume of MgO$_4$, 
as compared with ZnO$_4$. 
The bond valence parameter of Mg$-$O ($R_0=0.693\,\AA$) is smaller than 
that of Zn$-$O ($R_0=0.704\,\AA$), implying that in general Mg$-$O bonds 
are shorter than Zn$-$O bonds, for the same coordination numbers. 
Similarly, the 4-coordinate ionic radius of Mg$^{2+}$ (0.57\,\AA) is
smaller than  that of Zn$^{2+}$ (0.60\,\AA).  In order for the MgO$_4$
tetrahedron to have a smaller volume than ZnO$_4$, either the
tetrahedral height or the base area should shrink, or an
isotropic volume change could occur.  From the PDF analysis, the $c$ parameter 
around Mg showed a clear decrease, whereas the $a$ parameter appears to be
nearly unchanged. The $c$-axial parameter contraction of MgO$_4$ unit can be 
deduced also from the composition-dependent changes of the average structure 
of \ZM. 

\begin{figure}  \smallskip 
\centering \includegraphics[width=8cm]{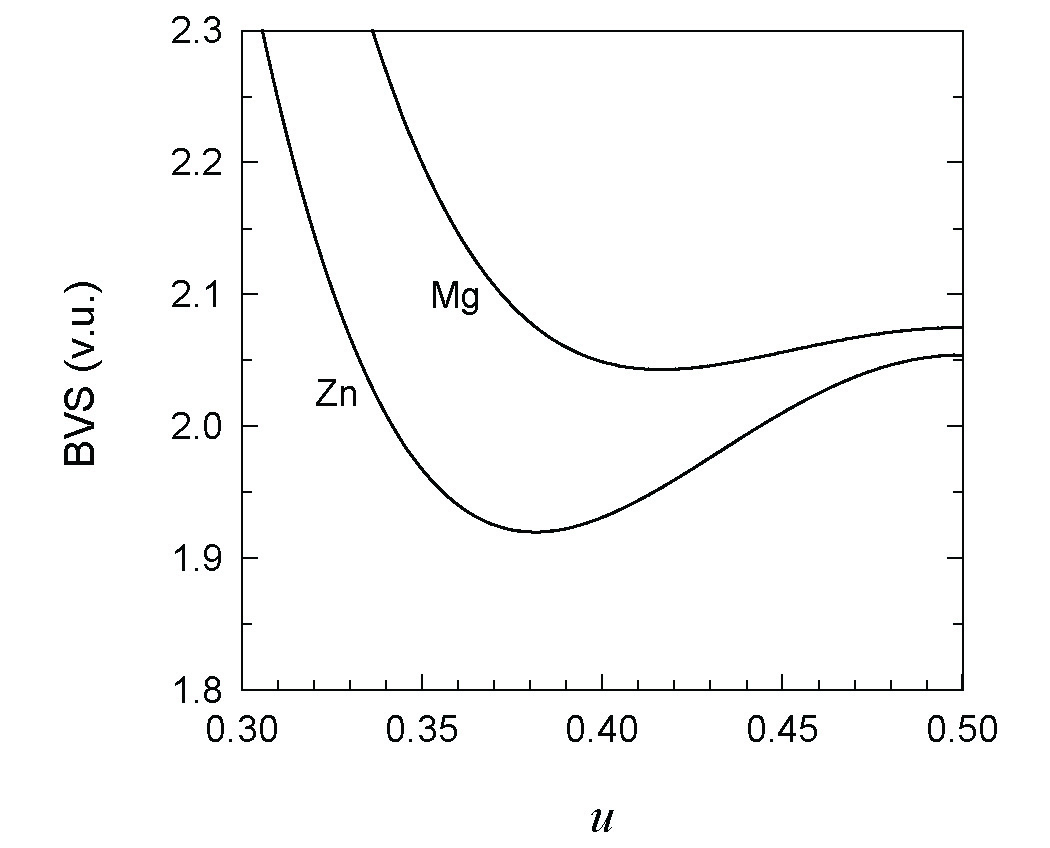} 
\caption{$u$-dependent bond valence sums for Mg and Zn in the tetrahedral 
spaces of their oxides. The dimensions of MgO$_4$ and ZnO$_4$ tetrahedra are from 
the neutron PDF refinements of \ZMM\/ and ZnO, respectively. 
For $u$ near 0.5, a fifth oxygen atom was added to the coordination shell 
to account for the pseudo-bipyramidal configuration.}
\label{fig:calcbvs}  \end{figure}

The variable $u$ is the only position parameter in the wurtzite structure. 
It corresponds to the ratio of the apical bond distance to the $c$ length, 
$d_\mathrm{apical}/c$, or $d_\mathrm{apical}/2h$, where $h$ is 
the height of the tetrahedron. 
For \ZMM, $u$(Mg) is determined to be 0.395, a value that finds 
no correspondence to the experimental structures of pure wurtzites. 
Existing wurtzites have $u$ values within the range of 0.374 to 0.383. 
Such a high $u$ found for MgO$_4$ is closely linked with 
its extremely small aspect ratio ($2h/a=1.509$, as compared to 
the ideal value 1.633). 
To compare the potential fields within the tetrahedral spaces of MgO$_4$ 
and ZnO$_4$, the valence sums of Mg and Zn were calculated as functions 
of $u$ (Fig.\,\ref{fig:calcbvs}). 
In ZnO, the tetrahedral cavity is slightly oversized for Zn, and the BVS(Zn)
has a minimum of $\approx$1.9\,v.u. at $u\approx$ 0.382. The potential
minimum suggested by BVS(Zn)  is in a good agreement with the
experimental $u$ for ZnO. 
However, the MgO$_4$ shell in \ZM\/ is 
rather small for Mg, resulting in a global overbonding situation. 
The BVS(Mg) is minimized at a markedly higher $u$ range than for 
BVS(Zn), making it clear why the experimental $u$(Mg) 
is conspicuously larger than $u$(Zn). 

\begin{figure}  \smallskip \centering 
\centering \includegraphics[width=8cm]{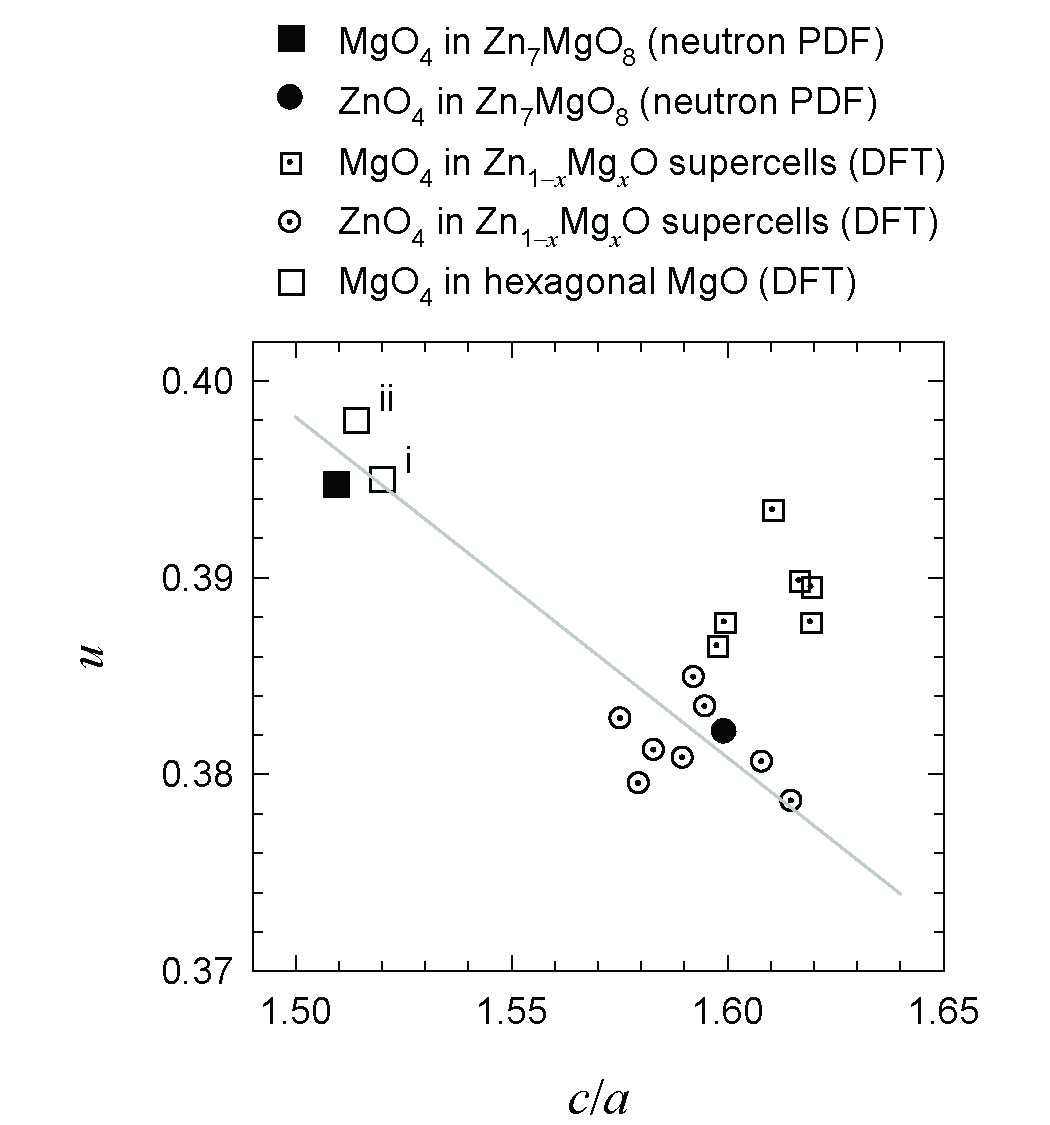} 
\caption{Aspect ratio $c/a$ and internal parameter $u$ for 
MgO$_4$ (filled squares) and ZnO$_4$ (filled circles) tetrahedra 
in \ZMM, as analyzed by neutron PDF refinements. 
DFT results for wurtzite MgO and \ZM\/ supercells are compared 
together: Open squares correspond to the MgO$_4$ geometry in 
hexagonal MgO (i: ref.\cite{Janotti} and ii: ref.\cite{Gopal}), 
and the dotted open squares and dotted open circles represent 
the MgO$_4$ and ZnO$_4$ geometries found in 
\ZM\/ supercells\cite{Mala2}, respectively. 
The gray line defines the border where the axial cation-anion
bond along $c$ and three equatorial bonds become equidistant.}
\label{fig:MgO4} 
\end{figure}

Several authors have used first-principles density functional theory (DFT) 
to predict the relaxed structure of hypothetical wurtzite MgO.\cite{Limp,Gopal,Janotti} 
An earlier DFT study attempted to relax the energy-minimized wurtzite 
MgO structure, but only to obtain a hexagonal structure corresponding to 
the limit of $c$-axis compression ($c/a=1.20$, $u=0.5$).\cite{Limp} 
However in later studies, the energy minimizations for wurtzite MgO 
were achieved within appropriate boundary conditions; 
$c/a=1.514$, $u=0.398$ by Janotti $et\,al.$\cite{Janotti}, and 
$c/a=1.520$, $u=0.395$ by Gopal and Spaldin\cite{Gopal}. 
Therefore, wurtzite MgO, if it ever occurs, is expected to have 
an abnormally small $c/a$ ratio and large $u$ value, compared with 
common wurtzites including ZnO. 
Interestingly, the above computational approximations of the hexagonal MgO 
structure agree well with our PDF analyses of \ZM, which underscores 
that the MgO$_4$ fragment has smaller $c/a$ and larger $u$ than ZnO$_4$. 
Malashevich and Vanderbilt\cite{Mala1} and Fan $et\,al.$\cite{Fan} 
performed DFT computations of \ZM\/ supercell models, 
also deriving consistent conclusions that $c/a$ decreases with 
Mg-substitution, and that $u$(Mg) $>$ $u$(Zn).  
The results from the neutron PDF refinement and the above DFT computations 
are plotted together in Fig.\,\ref{fig:MgO4}, in which the distinct 
internal/external geometries of MgO$_4$ and ZnO$_4$ are contrasted. 
As a general trend, isolated MgO$_4$ tetrahedra always have larger 
$u$ parameters than ZnO$_4$ units. The extremely good agreement between 
the \textit{local} coordination of Mg in \ZMM\/ determined here from PDF 
studies and DFT calculations on end-member wurtzite MgO is remarkable. 
This is in part attributed to our use of the ratio $c/a$ as a comparison 
parameter. This ratio has fewer systematic errors than do the values 
of the lattice parameters taken separately. 

The total polarization $P_\mathrm{s}$ of polar oxides can be 
broken down into electronic, ionic, and piezoelectric contributions, 
related respectively to the polarizability of component ions, 
ionic arrangements, and lattice parameters. 
Mg-substitution decreases the electronic polarization 
($P_\mathrm{el}$) of ZnO, because of its greater ionic character, 
whereas the concomitant $c$-axis compression is expected to increase 
the piezoelectric polarization ($P_\mathrm{pz}$). 
Quantitative estimation of the above two contributions 
($P_\mathrm{el}$, $P_\mathrm{pz}$) requires sophisticated theoretical analyses 
and cannot be provided here for ZnO and \ZMM. 
Meanwhile, the ionic polarization ($P_\mathrm{ion}$) of \ZM\/ 
can be approximated by a simple electrostatic model 
using the $c$-axial positions of Zn, Mg, and O. 
We calculated $P_\mathrm{s}$ of \ZMM\/ and ZnO, 
assuming point charges ($q=$\,+2 for Zn and Mg, and $-$2 for O), 
and using the atomic coordinates as determined by neutron PDF 
analysis (see Appendix). 
Separate reference structures were used for ZnO and \ZMM, with 
the corresponding experimental $a$ and $c$ values, but with $u$ = 0.375, 
so that within isolated (Zn,Mg)O$_4$ moieties the net centers of the 
cation and anion charges coincide. 
The calculated $P_\mathrm{s}$ values (= $P_\mathrm{ion}$ here), along $c$, 
are $-5.18\,\muup\mathrm{C}/\mathrm{cm}^2$ for ZnO and 
$-5.25\,\muup\mathrm{C}/\mathrm{cm}^2$ for \ZMM\/ 
($\Delta P_\mathrm{ion}=-0.07\,\muup\mathrm{C}/\mathrm{cm}^2$). 
For comparison, Malashevich and Vanderbilt\cite{Mala1} have employed the 
Berry-phase approach, to obtain electric polarization of 
a Zn$_5$MgO$_6$ ($x=\frac16$) supercell. 
They find $\Delta P_\mathrm{ion}=-0.22\,\muup\mathrm{C}/\mathrm{cm}^2$,
$\Delta P_\mathrm{el}=+0.01\,\muup\mathrm{C}/\mathrm{cm}^2$, 
and $\Delta P_\mathrm{pz}=-0.81\,\muup\mathrm{C}/\mathrm{cm}^2$, 
relative to $P_\mathrm{s}(\mathrm{ZnO})=-3.22\,\muup\mathrm{C}/\mathrm{cm}^2$. 
In fact, the following relations hold true in general for other \ZM\/ 
supercells ($x=\frac14, \frac13, \frac12$) studied by them: 
$\Delta P_\mathrm{el}>0$, $\Delta P_\mathrm{ion}<0$, 
$\Delta P_\mathrm{pz}<0$, and 
\textbar$\Delta P_\mathrm{el}$\textbar\/ $\ll$ 
\textbar$\Delta P_\mathrm{ion}$\textbar\/ $<$ 
\textbar$\Delta P_\mathrm{pz}$\textbar.    
We therefore expect that the total polarization of \ZM\/ should 
be larger than that of ZnO, with the increases of piezoelectric 
and ionic polarizations dominating the decrease of the electronic 
contribution. 
We note that $P_\mathrm{s}$(ZnO) from Malashevich and Vanderbilt differs 
somewhat from the results of other DFT studies\cite{Gopal,Corso,Bernardini,Noel} 
which reported 
$P_\mathrm{s}\mathrm{(ZnO)}=-5.0\sim-5.7\,\muup\mathrm{C}/\mathrm{cm}^2$. 
But since we are primarily interested in the relative portions of 
$P_\mathrm{el}$, $P_\mathrm{ion}$, and $P_\mathrm{pz}$ in the $P_\mathrm{s}$, 
and their changes with respect to the Mg-substitution in ZnO, we ignore the 
small discrepancies in the absolute values. 

The result, an increase of $P_\mathrm{ion}$ upon Mg-substitution, 
differs from our previous estimates based solely on synchrotron x-ray
diffraction analyses.\cite{Kim1,Kim2} 
This inconsistency arises mostly from the improved models for the 
\ZM\/ supercell structures that are permitted by the superior neutron 
scattering results. 
Figure\,\ref{fig:SFG} clearly shows the greatly enhanced data quality of 
the neutron scattering measurements, compared with x-ray experiments. 
The neutron data lead to better precision 
(see Table\,\ref{tab:Riet} and ref.\cite{Kim1}), as well as better accuracy, 
in the subsequent structure analyses. 

It is also noteworthy that the structure modeling for the PDF 
or Rietveld analyses can confine the range of structural information that 
is sought. 
For example, Rietveld refinement of \ZM\/ used only one positional 
parameter, and therefore MgO$_4$ and ZnO$_4$ tetrahedra are characterized 
by the same $c/a$ and $u$. In the x-ray PDF refinement, two position 
parameters were refined, so that we could differentiate $u$(Mg) and $u$(Zn), 
but not their $c/a$ ratios. However, for the neutron PDF refinement, we used 
four positional parameters as variables, and were able to distinguish the 
MgO$_4$ and ZnO$_4$ geometries through both $c/a$ and $u$. 
Again, the choice of a structural model relies on the quality of the 
diffraction data, for this application on a system with relatively 
low-atomic-number elements. The neutron data showed sufficient integrity 
for testing various structure models, which were not possible for the 
x-ray data alone. 

\section{Conclusion}

Detailed structural analyses of \ZM\/ ($0\,\leqslant\,x\,\leqslant\,0.15$) 
solid solutions, focusing on the dissimilar local geometries of Mg and Zn, 
have been performed by using Raman scattering, $^{67}$Zn/$^{25}$Mg  
NMR spectroscopy, and neutron Rietveld/PDF analyses. 
Line broadenings of the Raman and $^{67}$Zn/$^{25}$Mg NMR spectra 
imply that Mg-substitution into a ZnO-rich lattice gradually increases 
the configurational disorder and crystal defects in the lattice. 
Rietveld refinement of time-of-flight neutron diffraction patterns 
and Raman peak profile analyses show that the macroscopic lattice 
distortions in \ZM\/ wurtzites develop in a way such that the $c/a$ ratio 
decreases with Mg content $x$. 
A real-space neutron PDF analysis using a supercell model of Zn$_7$MgO$_8$ 
reveals that the Mg and Zn atoms in \ZM\/ have markedly distinct 
local geometries: MgO$_4$ tetrahedra are smaller in height and volume 
and have a larger internal parameter $u$. 
The wurtzite structural parameters $c/a$ and $u$ for MgO$_4$ deviate
from their ideal values, in agreement with computational predictions of 
wurtzite MgO structures. 
Previous DFT studies on \ZM\/ supercells and the present neutron PDF 
analysis strongly suggest that Mg-substitution will 
increase the spontaneous polarization of ZnO. 

\section*{ACKNOWLEDGEMENTS}  The authors acknowledge support from 
the National Science Foundation through the MRSEC program (DMR05-20415) 
and from the Department of Energy, Basic Energy Sciences, 
Catalysis Science Grant No. DE-FG02-03ER15467. 
This work has benefited from the use of NPDF at the Lujan Center at 
Los Alamos Neutron Science Center, funded by DOE Office of Basic Energy 
Sciences. 
Los Alamos National Laboratory is operated by Los Alamos 
National Security LLC under DOE Contract DE-AC52-06NA25396. 
The authors are grateful to the NSF-supported National High Magnetic Field 
Laboratory, in Tallahassee, Florida for access to the high-field (19.6\,T) 
NMR facilities, and to Zhehong Gan for 
assistance with the NMR measurements. 
Andrei Malashevich and David Vanderbilt 
kindly provided the DFT optimized structural data for \ZM\/ supercells. 
The authors also thank Brent Melot and Daniel Shoemaker for the neutron 
data collection.

\section*{APPENDIX}
Complete list of atomic coordinates for the 16-atom wurtzite supercell Zn$_7$MgO$_8$ 
($a=b=6.5054\,\AA$, $c=5.2015\,\AA$, $\alphaup=\betaup=$ 90$^{\circ}$, 
$\gammaup=$ 120$^{\circ}$), 
as determined by the neutron PDF analysis. 
\begin{table}[b]
\begin{ruledtabular} 
\begin{tabular}{lllll} 
Atom & $x$ & $y$ & $z$ & Type \\
\hline
Zn  & 0.1667 & 0.3333 & 0 & Zn1 \\
    & 0.6667 & 0.3333 & 0 & Zn3 \\
    & 0.1667 & 0.8333 & 0 & Zn1 \\
    & 0.6667 & 0.8333 & 0 & Zn1 \\
    & 0.3333 & 0.1667 & 0.5 & Zn2 \\
    & 0.8333 & 0.1667 & 0.5 & Zn2 \\
    & 0.8333 & 0.6667 & 0.5 & Zn2 \\
Mg  & 0.3333 & 0.6667 & 0.486(8) & Mg \\
O   & 0.1667 & 0.3333 & 0.386(3) & O2 \\
    & 0.6667 & 0.3333 & 0.3825(8) & O3 \\
    & 0.1667 & 0.8333 & 0.386(3) & O2 \\
    & 0.6667 & 0.8333 & 0.386(3) & O2 \\
    & 0.3333 & 0.1667 & 0.8825(8) & O3 \\
    & 0.8333 & 0.1667 & 0.8825(8) & O3 \\
    & 0.3333 & 0.6667 & 0.858(5) & O1 \\
    & 0.8333 & 0.6667 & 0.8825(8) & O3 \\ 
\end{tabular} 
\end{ruledtabular} 
\end{table} 

\end{document}